\documentstyle[12pt]{article}
\let\ssection=\section
\renewcommand{\section}{\setcounter{equation}{0}\ssection}

\begin{document}
\begin{titlepage}
\hspace{9truecm}Imperial/TP/98--99/23

\begin{center}
{\large\bf On the Emergence of Time in Quantum
Gravity\footnote{To appear in {\em The Arguments of Time},
ed. J.~Butterfield, Oxford University Press, 1999.}}
\end{center}
\vspace{0.8 truecm}

\begin{center}
            J.~Butterfield\footnote{email:
            jb56@cus.cam.ac.uk;
            jeremy.butterfield@all-souls.oxford.ac.uk}\\[10pt]
            All Souls College\\
            Oxford OX1 4AL
\end{center}
\begin{center}
and
\end{center}
\begin{center}
            C.J.~Isham\footnote{email: c.isham@ic.ac.uk}\\[10pt]
                The Blackett Laboratory\\
            Imperial College of Science, Technology \& Medicine\\
            South Kensington\\
            London SW7 2BZ\\
\end{center}
\begin{center}  13 December 1998 \end{center}

\begin{abstract}
We discuss from a philosophical perspective the way in
which the normal concept of time might be said to `emerge'
in a quantum theory of gravity. After an introduction, we
briefly discuss the notion of emergence, without regard to
time (Section 2). We then introduce the search for a
quantum theory of gravity (Section 3); and review some
general interpretative issues about space, time and matter
(Section 4). We then discuss the emergence of time in
simple quantum geometrodynamics, and in the Euclidean
approach (Section 5). Section 6 concludes.

\end{abstract} \end{titlepage}
\section{Introduction}
\label{Sec:Intro} The discovery of a satisfactory quantum
theory of gravity has been widely regarded as the Holy
Grail of theoretical physics for some forty years. In this
essay, we will discuss a philosophical aspect of the search
for such a theory that bears on our understanding of time:
namely, the senses in which our standard ideas of time, and
more generally spacetime, might be not fundamental to
reality, but instead `emergent' as an approximately valid
concept on sufficiently large scales.

In taking up this topic, our aim in part is to advertise to
philosophers of time an unexplored area that promises to be
fruitful. That is: we maintain that beneath the sometimes
rebarbative technicalities of the research programmes in
quantum gravity, there are conceptual topics that are
sufficiently related to current philosophical debates, that
philosophers can both contribute to those topics, and have
some light shed on their own debates. More specifically, we
believe some general philosophical claims about emergence,
which we will state in Section \ref{Sec:RSE}, are well
illustrated by some of these research programmes.
Furthermore, by addressing this topic, and illustrating it
with these programmes, our discussion will complement the
essays by Barbour and Kucha\v{r} (this volume).

We need first to introduce our topic: (this will also yield
a prospectus of the later Sections). The first thing to
note is an ambiguity in `emergence' and cognate words. In
everyday language, `emergence' suggests a process in time,
as in `the woman emerged from the crowd', or even an
evolutionary process, as in `amphibians emerged from
fishes'. But in this essay, we intend `emergence' in the
non-temporal, philosophical sense of, roughly, reduction,
or more often, some relation similar to (usually weaker
than) reduction; as in `the mental properties of a person
emerge from their physical properties'; or `the laws of
chemistry emerge from the basic equations of physics'.

We shall discuss the notion of emergence (and reduction) in
more detail in Section \ref{Sec:RSE}. But we can already
state the broad idea of the emergence of time. It is that
the concept of time within present physical theories---by
which we mean Newtonian physics, special relativity, and
general relativity---is an approximation within a quite
different conceptual framework that is associated with a
quite different theory. Arguably, our present theories'
concept of time stems from our psychological awareness of
the temporal ordering of events---a property that is
reflected mathematically in physics by the use of the real
numbers (a totally ordered system)\footnote{Those who have
succumbed to the seductive allure of the archetype of
`eternal recurrence'---for example, in our own time:
Nietzsche and T.S.~Eliot---would want to model points in
time by the points on a circle, rather than the real line.}
to represent `points' in time. In turn, these real numbers
are regarded as the `temporal coordinate' in a
four-dimensional differentiable manifold, whose remaining
three coordinates represent space. For Newtonian physics,
this distinction between `temporal' and `spatial'
coordinates is a fundamental ingredient in the
corresponding theories. In the case of relativity theory
(special or general), the distinction is coded in a metric
tensor: essentially, a mathematical structure that
specifies both spatial lengths and temporal intervals
between points in a unified spacetime.

In this context, the idea of the emergence of time is,
roughly speaking, that the concept of time as a continuum
ordering of events is only an approximation, valid for
sufficiently large scales of time and length (and,
correspondingly, sufficiently low energies\footnote{In a
quantum theory, the scales of time or length are related to
the inverse of energy via Planck's constant.}), to what is
expected to be a quite different concept in the, as yet,
unknown full quantum theory of gravity. Note that, for
brevity, we shall often talk of the `emergence of time',
not always adding `and of space' or `of spacetime'; but in
the light of relativity's unification of space and time, it
is reasonable to expect that both space and time are
emergent concepts.

It is already clear that a full discussion of the emergence
of time in quantum gravity would be a very daunting task.
It would involve probing an `intellectual space' with three
different `coordinates': {\em which\/} aspects of time are
emergent, in {\em which\/} sense, and in {\em which\/} of a
range of quantum gravity programmes. A full discussion
would be beyond our---and we dare say,
anybody's---competence, if all three coordinates are given
a wide range of values, which each get a detailed
treatment.

It is not just that among the putative aspects of time are
metaphysical matters such as the relation of time to
causation, and whether there is temporal becoming: matters
which are controversial, in themselves ({\em cf.\/} Tooley,
this volume) and in relation to present-day physical
theories, let alone some future theory of quantum gravity.
Also, even while confining ourselves to those aspects of
time that are uncontentious parts of our present theories'
treatment (as in the above paragraph), the other
`coordinates' are problematic. As we shall see in Section
\ref{Sec:RSE}, the general notion of emergence is vague,
and contentious. And once we settle on some such notion,
there is a wide and disparate range of quantum gravity
programmes to consider. Furthermore, most of these
programmes face hard conceptual problems (as well as
technical ones), that often involve how time is meant to
emerge. We shall mention some of these programmes and
problems in Section \ref{Sec:EnterpriseQG}; and in Section
\ref{Sec:stm}, we discuss some interpretative issues in
more detail.

Thereafter, we shall confine ourselves to a fragment of the
overall task. Namely, we will consider just two forms of
the canonical quantum gravity programme: (i) quantum
geometrodynamics, viewed as the general analysis of the
Wheeler-DeWitt equation (Section
\ref{Sec:CanQuGrav}---5.4); and (ii) the Euclidean
programme, which uses functional integrals of Euclidean
$4$-metrics to construct specific solutions of the
Wheeler-DeWitt equation (Section \ref{SubSec:ECG}).

These programmes are very closely related. Indeed, one
could equally well say, individuating `programmes' rather
differently, that the latter is part of the former.
Unsurprisingly, therefore, the latter faces much the same
conceptual problems as the former. It has, however,
distinctive features which merit the separate discussion in
Section \ref{SubSec:ECG}.

We should emphasise at the outset that these two programmes
are {\em not\/} now among the main current approaches to
quantum gravity. Yet there are three good reasons for
choosing them here: reasons relating to the present state
of the philosophical literature on quantum gravity, and to
this volume.

First: the emergence of time is related to the so-called
`problem of time', which has already been discussed by
philosophers (for example, \cite {BE99}); and though this
problem crops up in one form or another for all quantum
gravity programmes, it has been most studied, and is best
understood, for quantum geometrodynamics. Second, and
related: Kucha\v{r} (this volume) discusses precisely this
problem, for quantum geometrodynamics; so our choosing the
same programme gives us the opportunity to write a
complementary essay.

Our third reason relates to the Euclidean programme,
specifically to a model of quantum cosmogenesis made within
it by Hartle \& Hawking \cite{HH83}, viz.\ the `no-boundary
proposal'. A considerable philosophical, as well as popular
scientific, literature about the emergence of time has
grown up around this proposal. But we believe that much of
that literature is marred by not taking sufficient
cognizance of the problem of time, which besets the
Euclidean programme as much as quantum geometrodynamics. In
a nutshell: by ignoring the problem of time, much of the
literature makes it look as if the no-boundary proposal
treats the emergence of time as a process in time, {\em
i.e.}, according to the temporal sense of `emergence' noted
above. And then, unsurprisingly, the no-boundary proposal
seems to face conundrums or even contradictions. But these
are an artefact of ignoring the problem of time. In other
words: we agree that the no-boundary proposal is very
problematic: but this is because it must confront the
problem of time (to whose solution it makes no
contribution), not because of such conundrums---which we
believe rather facile.

\section{Emergence in General}\label{Sec:RSE}
\subsection{Prospectus}
\label{prospectus;emergce} In this Section we discuss
emergence in general, without regard to the idea of time
emerging. Although space is short, a brief discussion is
called for, for two reasons. First, philosophy has no
definite and established usage, for `emergence' and cognate
words. `Emergence' tends to be used as an
alternative---usually, a weaker alternative---to
`reduction': a concept, or a theory or some similar item
(such as a law or a model) is said to be, not reducible to
another such item, but merely emergent from it. But this
leaves open what exactly emergence is (not least because
`reduction' also has no definite and established meaning!),
and even what are the items between which it holds. So we
need to clarify the terms.

Second, and more important, we also want to make some
claims about emergence and its relation to reduction:
claims which will be illustrated by the quantum gravity
programmes discussed in Sections 4 and 5---but which we
think are worth stating in general. These claims will
follow from some criticisms (presented in Sections 2.2,
2.3) of a familiar, comparatively precise notion of
reduction, and an apparently weaker alternative,
supervenience.

To keep the discussion short, we will fix on one
traditional choice of the items between which relations of
reduction, supervenience and emergence hold. Namely:
theories, as understood in the syntactic conception (and in
formal logic), {\em i.e.}, understood as sets of sentences
closed under deduction. But we believe our arguments and
claims hold good, under other choices; in particular, on
the main rival conception of theories, viz.\ the semantic
conception according to which theories are classes of
models.

We will begin (Section 2.2) by introducing a precise notion
of reduction for theories. It is, in logicians' jargon,
`definitional extension'. Apart from the advantage of
precision, this notion has been immensely influential in
twentieth century philosophy: in the first place, in logic
and foundations of mathematics, and then in philosophy of
science. It has been the formal core of many proposed
analyses of the relation of reduction between scientific
theories. More precisely: on its own, definitional
extension is too weak---it applies in cases where
intuitively there is irreducibility. So analyses of
reduction typically add informal clauses, to try and
capture the idea that the reduced theory is `just a part'
of the reducing theory, as regards its concepts,
explanatory resources {\em etc}. We will maintain that such
considerations are so heterogeneous that there seems little
prospect of a general formal definition of reduction.

This notion of definitional extension will also lead us to
an apparently weaker, but also quite precise, notion:
supervenience (Section 2.3). Because supervenience is
apparently weaker, it has been a favoured candidate,
especially in recent metaphysics, for capturing the idea of
emergence, understood as a weaker alternative to reduction.
But we will point out that whether supervenience is really
weaker than definitional extension turns out to be a subtle
matter of logic. We will also claim that because
supervenience, like definitional extension, is a formal
notion, it is sometimes too weak. That is to say: it
applies in cases where intuitively a theory is {\em not\/}
emergent from another.

In Section 2.4, we spell out the upshot of this discussion
for the notion of emergence, as some weaker alternative to
reduction. Just as for reduction, the factors that prompt
us to talk of emergence are too informal, and vary too much
from one case to another, for there to be much prospect of
a satisfactory general, formal definition of emergence. And
being emergent is even compatible with reduction in our
initial, precise (and often too weak!) sense of
definitional extension. We will conclude that, rather than
seeking such a general definition, we need to bear in mind
the variety of ways that theories can be related: in
particular, with one theory being in some sense a limit of
the other, or an approximation to it.

\subsection{Reduction}
\label{subsec2ofemergce}
\paragraph{2.2.1 Introducing Definitional Extension}\label{para;DE}
The intuitive idea of one theory $T_1$ being reduced to
another $T_2$ is the idea of $T_1$ being shown to be a part
of $T_2$. The notion of definitional extension makes this
idea precise in two main ways. First, it focusses on the
syntactic conception of theories. This immediately gives a
notion of $T_1$ being a part of $T_2$, viz.\ when the
theorems of $T_1$ are a subset of those of $T_2$. (This is
called $T_1$ being a sub-theory of $T_2$.) However, one
needs to avoid confusion that can arise from the same
predicate (or other non-logical symbol) occurring in both
theories, but with different intended interpretations. This
is usually addressed by taking the theories to have
disjoint non-logical vocabularies. Then one defines $T_1$
to be a {\em definitional extension\/} of $T_2$, if and
only if one can add to $T_2$ a definition of each of the
non-logical symbols of $T_1$, in such a way that $T_1$
becomes a sub-theory of $T_2$. That is: In $T_2$, once
augmented with the definitions, we can prove every theorem
of $T_1$. (The definitions must of course be judiciously
chosen, with a view to securing the theorems of $T_1$.)

That is the principal idea of definitional extension. But
there is a second aspect, which concerns the question:
which operations for compounding predicates, and perhaps
other symbols, does one allow oneself in building the
definitions of the terms of $T_1$? It has been usual in
philosophy to consider a very meagre stock of operations,
viz.\ just the logical operations: the Boolean operations
of conjunction and negation, and the application of the
quantifiers `all' and `some'.

 Of course, this restriction does not just reflect the influence
of logic on analytic philosophy. It also has a deeper
rationale, relating to the striking success of studies in
logic and mathematics, from the mid-nineteenth century
onwards, in showing various pure mathematical theories to
be definitional extensions in the above sense, {\em i.e.},
using just these logical operations, of others. Indeed, by
concatenating such deductions with judiciously chosen
definitions, one shows in effect that all of classical pure
mathematics can in this sense be deduced from the theory of
sets. This remarkable result, showing how large an
expressive power can be obtained by applying this small
stock of logical operations to a small initial family of
predicates, is the source of the traditional philosophical
restriction to considering just these logical operations.

It is also of course a result whose impact on analytic
philosophy, including analytic philosophy of science,
cannot be over-estimated! For recall that the founding
fathers of analytic philosophy were very concerned with
applying the axiomatic method, and more generally the tools
of modern logic, to the natural sciences, especially
physics. The programme of stating physical theories
precisely, as theories in a formal language, and of
investigating their logical relations, preoccupied such
authors as Reichenbach and Schlick. More specifically, the
idea of one such theory being a sub-theory---{\em i.e.}, a
definitional extension---of another became the core idea in
analytic philosophers' proposals for the relation of
reduction between scientific theories.

We will turn to such proposals in a moment. But first, we
should note a general point about definitional extension:
for it will be needed in Section 2.3. It concerns the stock
of operations for compounding predicates {\em etc}.\ that
are used in building definitions. Namely: In discussing
physical, rather than pure mathematical, theories it is
very natural to augment the tiny stock of logical
operations, discussed above, with some of the standard
operations of mathematics, such as taking derivatives,
integrals, orthocomplements, completions {\em etc}. This
arises from the fact that even quite simple physical
theories such as, for example, Newtonian mechanics of point
particles interacting just by gravity, are complicated from
the viewpoint of logic and foundations of mathematics;
simply because such theories use mathematical apparatus
(for example, calculus) that is `high up' in the deductive
chain from basic logic and set theory. So when we undertake
to express such a theory in a formal language, we naturally
envisage adding the theory's distinctive physical
vocabulary, and axioms governing it, to a so-called
`underlying logic' that is much stronger than just basic
logic and set theory; {\em i.e.}, to one that contains
standard mathematical operations sufficient to yield the
mathematical apparatus; for example, calculus, which is
needed by the theory.

\paragraph{2.2.2 Definitional Extension in Philosophy of Science}
\label{para:defextensioninphilyscience} The best-known
proposed analyses of reduction, as a relation between
scientific theories, are by Hempel and Nagel, in various
papers from the 1940s to the 1970s. The standard reference
is \cite{Nag61} (Chapter 11):\footnote{For other
references, together with an evaluation of the proposals
for reduction in physics see, for example, \ Spector
\cite{Spe78}(Chapters 3-4, esp. p. 44).} Nagel adds to the
core idea of definitional extension some informal
conditions, mainly motivated by the idea that the reducing
theory should explain the reduced theory; and following
Hempel, he conceives explanation in deductive-nomological
terms. Thus he says, in effect, that $T_2$ reduces $T_1$
if, and only if:
\begin{enumerate}
\item $T_1$ is a definitional extension of $T_2$; and also

\item In each of the definitions of the terms of $T_1$
(which Nagel calls a `bridge law'), the definiens (in the
language of $T_2$) must play a role in $T_2$; so it cannot
be, for example, a heterogeneous disjunction.
\end{enumerate}
We need to make three comments on this sort of proposal;
(the details of Nagel's clause 2.\ will not concern us). In
effect, the first is a positive comment about definitional
extension. The second and third (in the next two
Paragraphs) are negative comments: in short, that
definitional extension is sometimes too weak for reduction,
and that it is sometimes too strong.

We said above that it was very remarkable that in effect
all of classical pure mathematics was a definitional
extension using just logical operations, of set theory. But
it is equally remarkable how many examples of definitional
extension there are in physics, at least once we give the
reducing physical theory a suitably strong underlying logic
({\em i.e.}, a rich enough set of operations) so as to
`extend its deductive reach'. Examples arise in almost any
case where there is a derivation of one physical theory
from another; and there are many such. For example: the
elementary theory of equilibrium classical statistical
mechanics for a strictly isolated system, which postulates
the microcanonical measure on the energy hypersurface, is a
definitional extension of the mechanics of the
micro-constituents, once we use an underlying logic strong
enough to express calculus (of many
variables).\footnote{The details, unimportant for the rest
of this paper, are as follows. The notion that this theory
adds to the mechanics of the micro-constituents is exactly
this microcanonical measure. But within the mechanics
together with an underlying logic containing calculus, we
can define both (i) the measure's domain of definition, and
(ii) the measure itself. (i) is just the (Borel subsets of)
the surface in phase space of given energy; and so is
definable. (ii) is defined by weighting the Lebesque
measure $\mu$ on phase space with the gradient $\nabla E$
of the energy E; and both $\mu$ and $\nabla E$ are
definable. Incidentally: for its empirical adequacy, the
theory needs to assume that the micro-constituents do not
interact; cf., for example, \cite{Khi49} (p. 36, 42-3,
111.)}

This point is worth stressing, since it tends to be
obscured by the traditional emphasis in the philosophical
literature on the apparent defects of definitional
extension as a general analysis, or part of a general
analysis, of reduction. Though we now turn to discuss those
defects, they should not obscure the remarkable number of
deductive interconnections between various pieces of
physics.

\paragraph{2.2.3 Definitional Extension is too Weak}
\label{para:defextensionistooweak} The point that
definitional extension is sometimes too weak for reduction
has been widely recognized.\footnote{But it is, we think,
not as widely recognized as the converse `Feyerabendian'
objection that definitional extension is too
strong---discussed in the next Paragraph. Smith
\cite{Smi92} (pp 30--32) is a laudable exception.} We can
express the point in general, as follows. Even if $T_1$ is
a definitional extension of $T_2$, so that (roughly
speaking) each of its predicates (or other non-logical
terms) is coextensive with a compound predicate built from
$T_2$'s vocabulary: there may well be aspects of $T_1$,
crucial to its functioning as a scientific theory, that are
not encompassed by (are not part of) the corresponding
aspects of $T_2$. For example, one might think that despite
the strict (exceptionless) coextensiveness of predicates,
$T_1$'s properties are different from those of $T_2$
(including $T_2$'s compound properties). Or one might think
that $T_1$ has aspects to do with explanation, or
modelling, or heuristics, that are not encompassed by $T_2$
(even taken together with the coextensiveness, given
suitable definitions, of predicates).

Clearly, what these aspects of $T_1$ that `outstrip' $T_2$
might be, varies from case to case; and being informal,
they are typically controversial. But both the general
point, and the controversial status of the details, is well
illustrated by Nagel's own proposed analysis of reduction,
and the subsequent literature. As we mentioned, Nagel's own
supplement to definitional extension, clause 2. above, was
motivated by the idea that $T_2$ should explain $T_1$,
where explanation is conceived in deductive-nomological
terms. And following Nagel, there were various rival
proposals about how to supplement definitional extension.
Some authors focussed on ontological issues about
property-identity, and thus on the definitions of $T_1$'s
terms (the bridge laws): for example, requiring them to be
statements of nomological coextensiveness. Other authors
focussed on whether reduction must include explanation, and
on what is required for an explanation; and so on.

Fortunately, we do not need to enter the controversy about
which such supplementary clauses might be right. We need
only note two points.

(A) The first relates to our statement in Paragraph 2.2.2
that physics provides many examples of definitional
extension. Namely: though this statement may seem
contentious, in the light of various doctrines about the
`disunity of physics', we now see that it is not. For such
doctrines say, in effect, that whichever supplementary
clauses are right, there are few pairs of theories
instantiating them---for physics is disunited. And that is
compatible with many instantiations of the `first clause',
definitional extension.

To put the same point in other words: physics abounds with
examples of definitional extension, provided that the
derived theory is conceived narrowly enough that it does
not have properties, explanatory resources {\em etc}.\ that
`ride free' of the deriving theory, and so block the
derivation. Indeed, our example of classical statistical
mechanics was deliberately chosen to illustrate this. Thus
we concede that:
\begin{enumerate}
\item Whether statistical
mechanics reduces to the mechanics of the individual
microscopic constituents is much debated (as is whether
thermodynamics reduces to statistical mechanics);

\item As regards our
example, one can take the view that the microcanonical
measure is a distinctively probabilistic concept, and so is
merely coextensive with its micromechanical {\em
definiens\/};

\item More generally, for the standard example of temperature:
one can take the view that although in elementary
statistical mechanics, the temperature of a sample of gas
is definable as the mean kinetic energy of the constituent
molecules, a more sophisticated treatment shows that
temperature indeed has a `distinctive identity' oustripping
this {\em definiens\/}. Its central role is to be a
parameter describing a probability distribution, and as
such it figures in the laws of statistical mechanics, and
its explanations of various phenomena, in many complex
ways: prompting the claim that statistical mechanics is
indeed not reducible to micro-mechanics.
\end{enumerate}
Our point is simply that these concessions in no way
undermine our example's claim of definitional extension.

(B) Our second point arising from the weakness of
definitional extension is simply that the ongoing
controversy over how to supplement definitional extension
in an analysis of reduction suggests that there may well be
no single `best' concept of reduction---no `essence' of
reduction to be winnowed out by analysis. The discussion of
the next paragraph will support this suggestion.

\paragraph{2.2.4 Definitional Extension is too Strong}
\label{para:defextensionistoostrong} We turn to the idea
that definitional extension, and thereby proposals like
Nagel's that incorporate it, is sometimes too {\em
strong\/} for reduction. Again, the idea is widely
recognized: Nagel's proposal, and variants that keep his
clause 1. (only adjusting his 2.), are often now regarded
as too strict. That is: there are intuitive cases of $T_2$
reducing $T_1$ without such a deduction of $T_1$. This sort
of objection goes back at least to Feyerabend \cite{Fey62},
who gives examples of $T_2$ reducing $T_1$ while being
inconsistent with it. For example, Feyerabend agrees with
Nagel that Newtonian gravitation theory reduces Galileo's
law of free fall; but he stresses that it is inconsistent
with the law, since it says (contrary to Galileo's law)
that the acceleration of a body increases as it falls
towards the Earth.

Along with many others, we agree with this objection, and
the general idea it prompts: that reduction often involves
approximation. Indeed, Nagel himself proposed that a case
in which $T_1$'s laws are a reasonable approximation to
what strictly follows from $T_2$ should count as reduction.
More generally, various authors have suggested, in the wake
of the critique by Feyerabend {\em et al.\/}, that
reduction often involves $T_2$ including some sort of
analogue, $T*$ say, of $T_1$. They require this analogue to
be close enough to $T_1$, in such matters as its
theoretical properties and the postulates concerning them,
and/or its explanatory resources, and/or its observational
consequences, that one is happy to say that `$T_2$ reduces
$T_1$'---rather than something indicating a greater `gap'
between them, such as `$T_1$ emerges from $T_2$', or even
`$T_2$ replaces $T_1$' or even `the theories are
incommensurable'.

Here our deliberate vagueness about the conditions required
of $T*$ is intended to signal both that the various authors
differ in their formulations, and that controversy
continues. Among the standard cases that get studied are
the pairs, Newtonian mechanics and special relativity, and
thermodynamics and statistical mechanics; (or better,
precise fragments of these very broad theories). While some
authors cite them as examples of reduction or emergence, on
account of such factors as the first theory providing an
approximation, or even some sort of mathematical limit, of
the second; other authors cite them as examples of
replacement or even incommensurability, on account of such
factors as the conceptual and explanatory disparities
between them.

Again, we fortunately do not need to enter into the
detailed formulations, or the controversy. We only need two
uncontroversial points arising from this discussion:
\begin{enumerate}
\item The fact that definitional extension often seems too strict
for reduction prompts one to appeal to such concepts as
approximation and limiting relations; and more generally,
to the reducing theory containing an analogue of the
reduced theory. So there are notions of reduction which at
least `come very close' to the idea of emergence.
\item The controversy suggests, as at the end of Paragraph 2.2.3, that
there may well be no single `best' concept of reduction;
and so no sharp division between reduction and such
concepts as `emergence', and even `replacement' and
`incommensurability'.
\end{enumerate}
So much for reduction, approached in terms of definitional
extension (using as the stock of compounding operations
either just the logical operations or some richer stock
including some standard mathematical operations). We turn
to supervenience.

\subsection{Supervenience}
\label{supervenience} Supervenience is a notion that is
apparently weaker than definitional extension, but also
quite precise. Because it is apparently weaker, it has been
a favoured candidate, especially in recent metaphysics, for
capturing the idea of emergence, understood as a weaker
alternative to reduction. Indeed, since supervenience is in
effect an infinitistic analogue of definitional extension,
supervenience promises to make the reduction/emergence
distinction as sharp as the distinction finite/infinite: or
at least it promises to do this, in so far as reduction
involves definitional extension. So at first sight,
supervenience promises to overcome the pessimism at the end
of the last Subsection, about sharply distinguishing
reduction and emergence.

But we suspect that this promise is illusory. After
presenting the idea of supervenience, we will point out
that whether supervenience is really weaker than
definitional extension turns out to be a subtle matter of
logic. We will also claim that because supervenience, like
definitional extension, is a formal notion, it is sometimes
too weak. That is: it applies in cases where intuitively a
theory is {\em not\/} emergent from another.

So we begin by presenting supervenience. It is normally
introduced as follows. One says that one family of
properties $F_1$ supervenes on another family $F_2$, with
respect to a given set $O$ of objects (on which both $F_1$
and $F_2$ are defined), if and only if any two objects in
$O$ that match for all properties in $F_2$ also match for
all properties in $F_1$. Or equivalently, one defines with
the contrapositive: any two objects that differ in a
property in $F_1$ must also differ in some property or
other in $F_2$.

However, this definition turns out to be equivalent to an
infinitistic analogue of definitional extension. That is,
it is equivalent to allowing each {\em definiens\/} used in
giving a definitional extension to be infinitely, rather
than finitely, long. The idea of allowing an infinite
sequence of compounding operations is at first sight
mind-stretching; and one worries that technical obstacles,
even paradoxes, may be lurking. But if one takes as one's
stock of operations, just the logical operations---the case
on which the logic and metaphysics literature have
concentrated---the idea is tractable, and well understood.
And it is indeed equivalent to supervenience as introduced
above, given natural ways of making each idea precise.
Accordingly, supervenience is sometimes called `infinitary
reduction', where `infinitary' means `infinite or finite';
and where `reduction' just connotes `definitional
extension'---no informal conditions like Nagel's 2.\ being
required.\footnote{Here is the idea of the equivalence:
suppose that supervenience as normally defined holds. Then
for each property $P$ in $F_1$, one can construct a
definition of it (as applied to $O$) by taking the
disjunction of the complete descriptions in terms of the
family $F_2$ of all the objects in $O$ that instantiate
$P$. This disjunction will indeed be infinite if there are
infinitely many ways objects can combine properties in
$F_2$ while possessing $P$. But in any case, supervenience
will ensure that objects matching in their complete
$F_2$-descriptions match as regards $P$, so that the
disjunctive {\em definiens\/} is indeed coextensive with
$P$.}

In view of this equivalence, metaphysicians have found
supervenience attractive for formulating doctrines of
emergence (though some might use another word than
`emergence' for the relation, similar to but weaker than
reduction, that they are trying to articulate). For
supervenience promises to sidestep controversial issues,
such as property-identity and explanation, that, as we saw
in Section 2.2, one has to address in order to get an
analysis of reduction.\footnote{A standard, relatively
uncontroversial example is given by paintings. Most agree
that their aesthetic properties (such as `is
well-composed') supervene on a suitably rich set of
non-evaluative pictorial properties (such as `is magenta in
extreme top-left corner'). And many who are happy to accept
supervenience have been sceptical that a property such as
`is well-composed' could be a finitely long compound of
pictorial properties---after all, they point out, we hardly
know how to begin writing such a definition, let alone how
to improve it and perfect it.}

However, we shall now argue that this advantage is often
illusory. There are two problems.

\paragraph{2.3.1 Is Supervenience Weaker?}\label{para:supervenienceweaker?}
The relation of supervenience to (finite) definitional
extension is more problematic than we have said---and than
most of the literature takes it to be! It turns out to be a
subtle matter of logic whether supervenience really is
weaker than definitional extension (and thereby an
attractive candidate for the notion of emergence). In fact
there are two points to be made here.

First, in the well-understood case where we consider only
the logical operations, it turns out that under wide
conditions (roughly: restriction to first-order languages),
supervenience is {\em equivalent\/} to definitional
extension! This is the content of Beth's theorem, which
says that for such languages, explicit definability ({\em
i.e.} definitional extension) and implicit definability
(roughly, our supervenience) are equivalent.\footnote{So
far as we know, the first paper to spell out this out, as a
threat to supervenience, is Hellman \& Thompson's paper on
physicalism \cite{Hel75}. They also suggest a reply to the
threat. They argue that we generally expect to cast
scientific theories in higher-order languages; for example,
in order to escape Godel's verdict of the incompletability
of first-order arithmetic. Since Beth's theorem fails in
general for such languages, supervenience may yet be weaker
than finite definitional extension in the philosophically
important cases: for example, for the formulation of
physicalism.}

Second, once we consider not only logical operations, but
also various standard operations of mathematics---as urged
at the end of Paragraph 2.2.1---the relation between
supervenience and definitional extension is much less
clear. And it is less clear in ways that threaten the idea
that supervenience is weaker. For on the one hand, there
are in general no well-defined limits to infinite
iterations of such operations. And on the other hand, we
noted at the end of Paragraph 2.2.2 the power of
definitional extension, once we admitted these mathematical
operations.\footnote{Besides, the fact that in a given case
one knows no finite definitional extension, and has little
idea how to go about constructing one, hardly show that
there is none: a finitely long definition might be so long,
say a million pages, as to be incomprehensible to the human
mind. Thus there might yet be a finite reduction of
paintings' aesthetic properties! For more discussion of the
`power of reduction', cf. \cite{Wil85}.}

\paragraph{2.3.2 Supervenience can be too Weak for Emergence}
\label{para:superveniencetooweak} The second problem is the
analogue for supervenience of the objection in Paragraph
2.2.3 to definitional extension, viz.\ that it is sometimes
too weak for reduction. That is: Paragraph 2.2.3 objected
that $T_1$ might be a definitional extension of $T_2$, yet
not be reducible to $T_2$, since it had properties,
explanatory resources {\em etc}.\ that outstripped those of
$T_2$. A corresponding point can be made here: $T_1$ might
supervene on $T_2$, and yet not be emergent from $T_2$.

To make this point precisely, we need to allow for the fact
that---setting aside, now, the problem of Paragraph
2.3.1---supervenience is logically weaker than (finite)
definitional extension. That is, we need to allow for the
fact that supervenience corresponds to there being
definitions of $T_1$'s predicates (or other terms) that are
each finite or infinite; it does not correspond to there
being at least one definition that is infinite. This last
one might well call `mere supervenience'; (or cumbersomely:
`supervenience, but not its special case, definitional
extension'). So the point is that $T_1$ might supervene on
$T_2$, and yet not be emergent from $T_2$---and this is
{\em not\/} just because $T_1$ is in fact a definitional
extension of $T_2$, and not emergent from $T_2$. In other
words: the point is that $T_1$ might `merely supervene' on
$T_2$, and yet not be emergent from $T_2$.

 How exactly this could happen will of course depend on one's
views about emergence. For example, if one takes emergence
(as against reduction) to require a certain sort of `gap'
between the properties of $T_1$ and $T_2$, then the point
will be that $T_1$ might merely supervene on $T_2$, without
there being such a gap. Similarly, if one instead takes
emergence (as against reduction) to require a certain sort
of `gap' between explanatory resources: $T_1$ merely
supervening on $T_2$ might not secure such a gap.

Admittedly, it is hard to give convincing examples of such
cases, for two reasons. First, as emphasised in Paragraph
2.3.1, real cases of mere supervenience are hard to come
by: for example, in first-order languages, supervenience
collapses into definitional extension. Second, it is a
vague and contentious matter what sort of `gap' indicates
emergence as against reduction. Indeed, as we saw in
Section 2.2: the controversy over the nature of reduction,
and especially over the informal clauses (about analogies,
limits and approximations between theories) that might
supplement or replace the formal notion of definitional
extension, suggests that there may be no single `best'
concept of reduction, and no sharp division between
reduction and emergence.

Furthermore, this second reason makes us sceptical of the
prospects of a general, formal analysis of emergence; (as
we were in Paragraph 2.2.3 and 2.2.4, about reduction). For
the factors that prompt people to say one theory is
emergent from, rather than reduced to, another---like the
`distinctive identity' of properties, the autonomy of
explanations {\em etc}.---are exactly the factors that
bedevil a general analysis of reduction. In other words,
the present problem---that supervenience threatens to be
too weak for emergence, because it is too formal to
encompass such factors---is likely to face other attempts
to capture the idea of emergence, using purely formal
methods. So it is a problem that will probably not be
sidestepped by changing formal notions.\footnote{For
example, we do not think it is sidestepped by formulating
supervenience in terms of the semantic conception of
theories; nor by replacing supervenience, by some single,
but general, limiting relation between the theories.}

\subsection{Emergence}
\label{SubSec:emerge} The upshot of this discussion is to
give us a heterogeneous picture of emergence. Though
`emergence of $T_1$ from $T_2$' connotes a greater gap
between the theories, than does `reduction', $T_1$ can be
emergent in various ways. Furthermore, being emergent is
compatible with our initial, formal (and often too weak!)
notion of definitional extension.

For the sake of Sections 5, we should fill in this picture
a bit: that is, give more detail about the variety of
relations, of limit and approximation, that can exist
between theories. For both limits and approximations, we
want to make two points; they are all illustrated by the
theories discussed in Section 5.
\begin{enumerate}
\item As to limits, both points are warnings: that technically
and conceptually, the situation can be much more subtle
than the elementary idea of a mathematical limit tends to
suggest. For first, a limit can be singular in the
mathematical sense: and when one goes to such a limit, for
some characteristic theoretical parameter of a physical
theory, the structure of the theory can change radically.
Indeed, the standard example of one theory having another
as a limit---special relativity having Newtonian spacetime
as its $c \rightarrow \infty$ limit---is a case of this:
the theoretical structure changes radically, with the
spacetime metric becoming degenerate.
\item Second, the relation between two theories often involves
more than one limiting process; and these will in general
not commute with one another. Again, a standard
example---the relation between thermodynamics and
statistical mechanics---provides cases. (For example,
Compagner \cite{Com89} discusses the relation between the
thermodynamic and continuum limits (among others) of
statistical mechanics.)

\item Turning to approximation, we should resist the temptation
to think that approximation is always directed at showing
that a theory is capable (given the validity of the
approximation) of predicting or explaining (`saving') some
observable phenomenon. For the aim is often instead to
illuminate the theoretical content of a complex (perhaps
computationally intractable) theory, without regard to
observational predictions, even in some very extended sense
of observation. (Both the temptation, and its being wrong,
are apparent from the discussion of $T*$ in Section 2.2.4:
we are tempted to think of the reduced theory $T_1$ as more
observational than the reducing theory $T_2$, so that $T*$
would be an approximation directed at securing
observations; but $T_1$ might be wholly theoretical, and
both it and its analogue $T*$ might be aimed primarily at
illuminating $T_2$.)

\item We should distinguish various ways in which one theory (or
piece of theory) can approximate another. Of course the
core idea is that values of physical quantities should be
close to each other (more exactly: in certain
circumstances, close enough for certain purposes!) But such
closeness can obtain in various ways. Here are three broad
(not exclusive) ways, which will crop up in Section 5.
\begin{itemize}
\item Neglecting some quantities, and arguing that such
neglect is justified, for example by analysing the
stability under perturbation of one's calculation, is a
well-nigh universal practice in physical theorizing. If the
neglect is indeed justified, the full theory is
approximated by the fragment that neglects some quantities.
(An important special case of this is neglecting quantities
describing components of some composite system, in favour
of collective quantities describing the whole.)

\item Instead of selecting some quantities and
neglecting others, one might instead select some states.
That is: there might be a subset of one theory's
state-space, in which each state approximates a state of
another theory.

\item Very often, these first two ways are combined:
a subset of states approximates states of the other theory,
{\em only\/} for some relevant quantities. Typically, such
an approximation is defined by an inequality (often a `much
greater than') between complicated quantities, which is to
be read as specifying those states that make it true.
\end{itemize}
\end{enumerate}

\section{The Enterprise of Quantum Gravity}
\label{Sec:EnterpriseQG} In order to discuss the emergence
of time in quantum gravity, we first need a brief review of
what the enterprise of quantum gravity involves. That is
the task of this Section; we will end by stressing some
ways that our topic differs from emergence elsewhere in
philosophy. This will lead in to a more detailed discussion
of some interpretative issues about space, time and matter
in Section 4.

\subsection{The Planck Scale}\label{Planck}
It should be admitted from the outset, that the subject of
quantum gravity is exceptionally difficult and problematic,
both in regard to mathematical and to conceptual issues.
Furthermore, these problems are compounded by the almost
complete lack of any unequivocal data to guide attempts to
construct such a theory. In this sense, twentieth-century
physics is a victim of its own success: the empirical
success of general relativity and quantum theory in their
present forms means that we lack data bearing on how we
might reconcile them (or more generally, replace them).

Furthermore, it is very hard to see how we could get such
data. This is because simple dimensional arguments suggest
that quantum gravity effects should be important at about a
length-scale called the `Planck length' $L_P$ defined as
$L_P:=({G\hbar\over c^3})^{1/2}$; where $G$ is Newton's
constant of gravitation, which measures the strength of
gravity; $\hbar$ is Planck's constant, which is
characteristic of quantum theory; and $c$ is the speed of
light. The value of the Planck length is around
$10^{-33}\mbox{cm}$, which is truly miniscule: the
diameters of an atom, nucleus, proton and quark are,
respectively, about $10^{-8}$, $10^{-12}$, $10^{-13}$, and
$10^{-16}$ cm. So the Planck length is as many orders of
magnitude from the (upper limit for) the diameter of a
quark, as that diameter is from our familiar scale of a
centimetre!

The same point can also be made in regard to the other
so-called `Planck scales': the Planck time, the Planck
energy, and the Planck temperature. These characterize
physical regimes at which quantum gravity effects are
expected to be important, and---perhaps---where the
continuum picture of space and time breaks down. Like the
Planck length, their values are such that they are
empirically inaccessible except in the very early universe.
For example, the Planck time $L_T$ is about $10^{-42}$
seconds, and the Planck energy is $10^{19}$ GeV.

\subsection{Some Motivations for Studying Quantum Gravity}
The central task of quantum gravity is to reconcile two of
the main pillars of twentieth-century theoretical physics:
general relativity and quantum theory. However, these two
theories are disparate in significant ways, and they need
to be reconciled. More precisely:
\begin{enumerate}
\item General relativity is a classical theory of gravity,
but---unlike all other classical theories---it does not
work within the framework of a {\em fixed\/} space and
time, but is rather a theory {\em of\/} space and time.
Specifically, general relativity employs a continuous
spacetime manifold, endowed with a spacetime metric that
determines the spatial lengths and times elapsed along
curves in the manifold. From a physical perspective, the
curvature associated with this metric is postulated to
describe the gravitational field: in particular, its value
at any point is dependent on the state of the (classical)
matter at that point. This relation is epitomised in the
famous Einstein field equations. (Special relativity also
postulates a continuous spacetime manifold, but the metric
is a special case with zero curvature.)

\item Quantum `theory' is not a theory {\em per se\/} but
is rather a certain mathematical and conceptual framework
within which specific theories are constructed. Typically,
these deal with various kinds of matter and the forces they
exert on each other, including the fundamental forces other
than gravity: the electromagnetic, weak and strong forces.
Quantum theories are conceptually very different from
classical theories: in particular, the state of a system
does not assign real-number values to physical quantities
but only probabilities for obtaining such values if
appropriate measurements are made. And, though they share
with general relativity the assumption that it is
appropriate to coordinatise space and time with real
numbers, they do not treat gravity as curvature of the
corresponding spaces: they use either Newtonian spacetime
(which corresponds to a degenerate spacetime metric), or
the flat metric of special relativity.
\end{enumerate}
The fact that general relativity treats matter classically,
and gravity as curvature, while our best theories of matter
are quantum theories using a flat metric, is enough to show
that some sort of reconciliation is needed. But there are
also several other, more specific, motivations for seeking
a quantum theory of gravity. For example:
\begin{itemize}
\item[(i)]
Gravity is, at least {\em prima facie\/}, a force like the
others, and these {\em have\/} turned out to be quantum in
nature.

\item[(ii)]
There has been considerable success in developing unified,
quantum-theoretical descriptions of the non-gravitational
forces (the most successful example being the unified
theory of the electromagnetic and weak interactions), which
naturally prompts the idea that gravity should be unified
with them.

\item[(iii)]
In the 1960's, spacetime singularities---at which the
values of physical fields like curvature and matter density
become infinite---were proven to be endemic to general
relativity. It has long been conjectured that quantum
gravity could remove such defects in the classical theory;
rather as the intrinsic instability of the classical atom
is resolved by the introduction of quantum theory. The main
justification for such an optimistic conjecture is the
existence of the Planck length $L_P\simeq
10^{-33}\mbox{cms}$: the hope is that the `zero-length'
spacetime singularities will no longer arise because of
quantum gravity effects that set in at this length scale.

\item[(iv)]
{Quantum theory has various internal problems of its own,
and it has been suggested that these could be resolved if
general relativity is introduced in an appropriate way. A
long-standing example of such a problem is the infinities
that arise in relativistic quantum field theory from the
singular nature of quantum fields at a spacetime point;
again, it is the existence of the Planck length that
encourages such speculations.

Another famous difficulty in quantum theory is the
notorious `measurement problem' with its associated idea of
a `reduction of the state vector'. There have been several
suggestions that reduction may be a real physical process
that is associated in some way with the gravitational
field. The key idea is that real reductions may arise for
`large' objects (thus avoiding the paradoxes associated
with Schr\"odinger's infamous cat); and the size of an
object is well reflected in the strength of gravitational
field it produces. }
\end{itemize}

These motivations are logically compatible, and one might
wish to endorse them all. But, in fact, when one tries to
develop a detailed theory of quantum gravity, it is
extremely hard to implement any of them in a coherent way;
and, in practice, the specific approach to quantum gravity
that is adopted often depends strongly on which motivations
are uppermost in the investigator's mind. Consequently, the
quantum gravity community holds many different views of
what such a theory should do, and many different approaches
have been tried.

At first glance, there may seem to be a reasonable
prognosis for the success of such an endeavour. For
example, there are various `quantization' procedures
whereby a quantum version of any given classical theory may
be constructed; and, therefore, one obvious approach to
quantum gravity, is to try to apply such a procedure to the
classical theory of general relativity. Another possibility
is to construct a quantum theory that is not a quantization
of general relativity, but which nevertheless has that
theory as a classical limit---superstring theory is a good
example of a structure of this type.

But despite much effort, using a variety of approaches and
heuristic ideas, it has turned out to be extremely
difficult to find a quantum theory of gravity. In
particular, the approaches based on techniques that have
been successful for the other forces, run aground.

Of course, we cannot give even a brief review of all the
attempts, and the remaining live options\footnote{Our paper
\cite{BI99} reviews some of these, from a non-technical
viewpoint; a more technical review is \cite{Ish97}.}, and
in this Section, we will confine ourselves to very brief
treatments of a few topics. First, we describe two of the
main current approaches to quantum gravity. Second, and
most relevant to our theme of the emergence of time: we
describe how these approaches assume a manifold structure
for spacetime, and how this can be questioned. (What this
assumption involves will be further discussed in Section
4.) Third, we take note of some ways in which the emergence
of time is expected to be very unlike other cases of
emergence.

\subsection{The Two Main Approaches}\label{twomain}
There are currently two main approaches to quantum gravity,
each of which has several variants and more specific forms.
The first exemplifies the strategy mentioned above, of
starting with a classical theory and quantizing it; the
second affords an example of moving in the opposite
direction. The two approaches are as follows:

\paragraph{The canonical approach:}
The basic idea here is first to recast Einstein's general
relativity in Hamiltonian form, in which it describes the
evolution in time of 3-dimensional geometries, {\em i.e.},
the geometry of an instantaneous 3-dimensional hypersurface
(a spacelike `slice') of the spacetime manifold (in
general, interacting with matter fields). The ensuing
Hamiltonian theory is then quantized by adapting to general
relativity one or other of the standard quantization
procedures that has been successful in quantizing some of
the other fundamental forces.

Nowadays, this approach is mostly pursued in a different
form, based on ideas of Ashtekar. The idea of `splitting'
spacetime into 3-dimensional slices, and conceiving
dynamics as evolution from one slice to another, remains;
but the basic dynamical variable is now, not a 3-geometry,
but a 3-connection (roughly: a notion of parallelism for
lines). This approach has lead to some radically new
perspectives on quantum geometry, not least via choosing
the basic variables to be loop-integrals of the
3-connection. However, for reasons given in Section
\ref{Sec:Intro}, this essay will focus on the older form of
the approach.

\paragraph{The superstring programme:}
Here, instead of quantizing general relativity, one starts
with a very different theory whose basic items are
quantized, 1-dimensional strings (or higher-dimensional
extended objects) propagating in a continuum spacetime.
There is a certain sense in which normal general relativity
emerges as a low-energy limit of this theory; and the hope
is that the other fundamental forces of nature are also
contained implicitly in the structure. Again, for the
reasons given in Section \ref{Sec:Intro}, we shall not
discuss this important programme further in the present
essay.

\subsection{Should We Assume a Manifold?}
\label{SubSec:Assumemanifold} We turn now to the issue of
the extent to which it is appropriate in quantum gravity to
assume the same `continuum' ({\em i.e.} manifold) structure
for spacetime as that employed in the classical theory of
general relativity. The first point to make is that, at
least in their current forms, both the above approaches to
quantum gravity do indeed assume a manifold. However, in
considering this issue in general it is helpful to think of
a `chain' of successively richer structure, that can be
added to a `bare' set of points, to yield the full
classical structure of a manifold with a spacetime metric.
For example, if $g$ is a Lorentzian metric on a spacetime
manifold $\cal M$, the pair $({\cal M},g)$ fits naturally
into the chain
\begin{equation}
\mbox{set of spacetime
points}\rightarrow\mbox{topology}\rightarrow
\mbox{differential structure}\rightarrow ({\cal M},g)
                \nonumber
\end{equation}
where the lowest level ({\em i.e.}, the left hand end) is a
set $\cal M$ of bare spacetime points (with the cardinality
of the continuum), which is then given the structure of a
topological space, which in turn is given the structure of
a differentiable manifold (only possible---of course---for
very special topologies) which is then equipped with a
Lorentzian metric to give the final pair $({\cal M},g)$.
Note that a variety of intermediate stages can be inserted:
for example, the link `$\mbox{differential
structure}\rightarrow({\cal M},g)$' could be factorized as
\begin{equation}
\mbox{differential structure}\rightarrow
    \mbox{causal structure}\rightarrow({\cal M},g).
                \nonumber
\end{equation}

Given such a chain, one can often distinguish approaches to
quantum gravity (or their specific forms) by `how far along
towards the right' they assume classical spacetime
structure. And one can distinguish approaches that match on
this question, by differences in their detailed assumptions
about this classical structure, or in their treatments of
the `right end of the chain': is it a quantization of the
given classical structure, or something else again?

Let us briefly illustrate this sort of classification with
a comment about each of the two main current approaches;
(there will be further illustrations in Sections 4 and 5.).
The approaches match on the initial question---they both
assume a manifold---but they differ on some of the basic
features of this manifold, such as its global topological
structure and dimension. In a bit more detail:
\begin{itemize}
\item
In the canonical approach to quantum gravity, a manifold
structure is assumed for 3-dimensional physical space.
Furthermore, in the classical formalism with which one
starts, a manifold structure is also assumed for spacetime,
which is taken, topologically, to be the product of a
(spatial) 3-manifold and the real line, representing time.

\item
In the original, perturbative, approach to superstring
theory, the oscillating strings are regarded as propagating
in a background spacetime, which also is taken to be a
manifold, albeit with a dimension that is possibly greater
than four (the extra dimensions are thought to be `rolled
up', like the lateral dimension of a cylinder whose
diameter is of Planck size, and hence go unnoticed).
\end{itemize}

This use of a continuum model for spacetime in quantum
gravity has been questioned. Some more radical approaches
propose that at sufficiently tiny length-scales
(sufficiently high energies) the postulate of a manifold
breaks down, and we need a quite different structure. In
particular, much effort is currently being devoted to
finding a non-perturbative form of string theory (the
so-called `$M$-theory') which is generally expected to
involve a novel mathematical model for spacetime: for
example, there are tantalizing hints that `non-commutative'
geometry may play a central role.

The idea that simple manifold concepts may not apply at
small distances was anticipated by Riemann in his famously
prophetic {\em Habilitationschrift\/} of 1854:
\begin{quote}
``Now it seems that the empirical notions on which the
metrical determinations of space are founded, the notion of
a solid body and of a ray of light, cease to be valid for
the infinitely small. We are therefore quite at liberty to
suppose that the metric relations of space in the
infinitely small do not conform to the hypotheses of
geometry; and we ought in fact to suppose it, if we can
thereby obtain a simpler explanation of phenomena.''
[Translated by Clifford, \cite{Cli73}]
\end{quote}
Here then is a more radical sort of sense in which time, or
better spacetime, might emerge. Of course, this radicalism
comes with a price. The usual tools of mathematical physics
depend so strongly on the real-number continuum, and its
generalizations (from elementary calculus `upwards' to
manifolds and beyond), that it is probably even harder to
guess what non-continuum structure is needed by such
radical approaches, than to guess what novel structures of
dimension, metric {\em etc}.\ are needed by the more
conservative approaches that retain manifolds. Indeed,
there is a more general point: space and time are such
crucial categories for thinking about, and describing, the
empirical world, that it is bound to be ferociously
difficult to understand their emerging, or even some
aspects of them emerging, from `something else'. (We
develop this point slightly in Section 3.5.)

However, we are sympathetic to this radical sort of
approach. And there are a number of proposals of this kind.
Given the preceding discussion, perhaps the clearest
examples are the suggestions to quantize structures that
lie to the left of manifold-structure, in the chain above.
For example, there has been work on quantizing topological
structure \cite{Ish89a}, and on the quantization of causal
sets \cite{Sor83,Sor91b}.

On the other hand, it would be precipitate to blame the
difficulties into which the conservative approaches ({\em
i.e.}, assuming manifolds) run, on their use of manifolds.
For first, precisely because they {\em are\/} conservative,
it is easier to develop them and to compare them with
existing theories, and thereby to discover their
limitations. And second, as we said above, to a significant
extent the difficulties arise from the inaccessibility of
the Planck scale (at least, in regard to terrestrial
experiments), and the associated lack of any specific data
to guide theoretical speculations. This leads to our fourth
topic.

\subsection{A Philosophical Warning}
\label{SubSec:philwarning} This inaccessibility not only
implies a lack of data. It is {\em so\/} extreme that it
suggests that those aspects of reality that require a
theory of quantum gravity for their description hardly
deserve such names as `appearance', `phenomenon', or even
perhaps `empirical reality'. Here of course our topic---the
emergence of time---touches on deep, in part Kantian,
themes about empirical knowledge and belief in general:
themes which we cannot properly take up here. But we should
take note of the Kantian position that it is not merely
very difficult to dispense with space and time in thought
about the empirical world (as we said above): it is
downright impossible.

Scientists are understandably sceptical of claims that
certain ways of thought are necessary, not least because
the history of science gives remarkable examples of the
creative, albeit fallible, forging of new concepts. But in
reply, it must here suffice to make three comments in
regard to the Kantian position. First, one might only
require that certain aspects of space and time, such as
their being continua, emerge; and a Kantian might allow
that these aspects are in principle dispensable (though
perhaps: in practice very hard to dispense with!). Second:
maybe the extreme inaccessibility just discussed---the
miniscule size of the Planck length and time {\em
etc}.---puts these aspects of reality ({\em i.e.}, those
requiring a theory of quantum gravity) outside the realm of
appearance in the Kantian sense.

Third, and more vaguely: we agree that our topic cannot
provide an `ordinary case-study' of emergence. And this is
not just because philosophical controversy continues about
how to understand one theory as emerging from another; nor
just because we are so far from having a satisfactory
theory of quantum gravity that one of the two theories
needed for a `case-study of emergence' is lacking. As noted
above, our predicament is aggravated by the continuum model
of spacetime being so crucial to our normal physical
description of the world. So we must be wary of assuming
that the emergence of time will be like more familiar cases
of emergence. In particular, we must bear in mind that
there is far from being a universal scientific agreement on
even the overall shape, let alone the details, of a
satisfactory theory of quantum gravity. Furthermore,
current approaches to quantum gravity face severe
conceptual, as well as mathematical, difficulties, and we
must be ready for a complex and unfamiliar relation between
the conceptual frameworks of our present theories---using
the standard notions of space and time---and those of the,
as yet, unknown theory of quantum gravity.

\section{Three Interpretative Issues}\label{Sec:stm}
In this Section, we discuss in more detail three
interpretative issues that were mentioned in Section
\ref{Sec:EnterpriseQG}, and to which the emergence of time
is closely tied. These topics are: the idea of a `fixed'
theoretical structure; the role of matter in understanding
spacetime; and the interpretation of quantum theory. We
will discuss them in order in the next three Subsections,
each leading in to the next; and we will illustrate the
first two topics with some classical spacetime theories
including general relativity.

\subsection{Fixity}\label{SubSec:Fixity}
Saying that a theoretical structure is `fixed' in a given
theory can have several different meanings, but they all
have the connotation that the structure is `given once for
all' in the theory. So to avoid confusion, one needs to
distinguish these meanings: in fact, we will spell out
three.\footnote{Another source of confusion is that there
is no established usage about which word to use. All of the
adjectives, `external', `prior', `absolute', `primitive'
and their cognates are sometimes employed. But for various
reasons, we prefer `fixed' and its cognates.} This is not
just a matter of clarifying terms: we shall see in Section
\ref{Sec:QGeom} that the problem of time---which, as noted
in Section \ref{Sec:Intro}, is entangled with the emergence
of time---arises essentially from the fact that in quantum
theory, time is treated as fixed (in all three of the
meanings below), while in general relativity it is not---it
is an aspect of the physical system.

In short, our three meanings of `fixed' are: being
classical; being non-dynamical; and being the same in all
models of the theory. As is suggested by the words
`classical' and `non-dynamical', these meanings will be
rather vague. But we will at least distinguish them
sufficiently to see that, with one exception, they are
mutually logically independent; and that is enough for our
present purposes.

The first meaning is common in discussions of quantum
gravity. Here, `fixed' is used to indicate that a structure
present in a classical theory is not quantized (in the
given approach to quantum gravity). This is especially
common in discussions of programmes that proceed by
quantizing a classical theory; (recall the discussion of
the chain in Section \ref{SubSec:Assumemanifold}).

Second, `fixed' is used to mean `not subject to dynamical
evolution' (often called, for short: non-dynamical). This
usage is not peculiar to quantum gravity, but occurs in a
variety of discussions of spacetime structure. In this
sense, the metrical structure of spacetime is fixed in
Newtonian physics and in classical and quantum theories
that work within a special relativistic (Minkowski)
spacetime. But it is not fixed in general relativity.

This second meaning is clearly logically independent of the
first. The case of quantum field theory in a curved
spacetime shows that the first meaning of `fixed'---not
quantized---does not imply the second, viz.\ being
non-dynamical: for example, the (unquantized) metric tensor
might obey a form of Einstein's equations in which the
right hand side is an expectation value of the
energy-momentum tensor of the quantized matter fields. And
the converse is also false: a structure can be
non-dynamical but quantized. An obvious example is a
conserved quantity in standard quantum theory: for example,
the total angular momentum in a system (like the hydrogen
atom) whose Hamiltonian is spherically symmetric. It is
clearly non-dynamical (in the sense that its eigenvectors
do not evolve in time; or, equivalently, the corresponding
operator in the Heisenberg picture does not change in
time.) but it is not classical since it is represented by
an operator $\hat{J}\cdot\hat{J}$, and most states do not
give it a definite value.\footnote{Of course, people also
use `classical' to mean a wider notion than `all quantities
having a definite value in all (pure) states'. For example,
they say it of any mathematical structure that is used in a
classical theory; so that, for example, the assumption of a
product manifold structure for spacetime by
geometrodynamics (Section 3.3) counts as classical. (And
they even say `fixed' to indicate this wider sense of
`classical'.) But even in this wider sense of `classical',
being non-dynamical does not imply being classical: again,
the angular-momentum operator $\hat{J}\cdot\hat{J}$ in
elementary quantum theory is an example.}

The third meaning of `fixity' is that a theoretical
structure is `fixed' if it is `given once and for all' in
the very formulation of the theory, {\em i.e.\/} it is the
same in all the different descriptions that the theory
allows of its subject matter; such structure is often said
to be part of the fixed `background'. So the idea assumes a
contrast between (i) the theory, that describes all the
patterns of behaviour of its subject-matter that are
possible according to the theory; and (ii) an individual
pattern. An individual pattern is typically a history of a
system of the kind described by the theory, and is called a
model; and the theory's main content is a set of dynamical
laws (laws of evolution over time)---in physical theories,
usually differential equations. So a fixed theoretical
structure is common to all such models.

One naturally expects that it would be easier to make this
third meaning precise than the first two, about being
classical and being non-dynamical. But in fact it is
surprisingly hard to make precise sense of a theoretical
structure being the same in all models of the theory. It
turns out that even when one uses a formal notion of model,
and so can precisely define associated notions such as
isomorphisms of models, it is difficult to define `fixed'
in such a way as to capture intuitions about what counts as
fixed---even in a small handful of precise and
well-understood theories, such as classical theories of
electromagnetism and gravity in Newtonian and relativistic
spacetimes. In fact, controversy continues about
this.\footnote{The word `absolute' is commonly used in
place of our `fixed'; a recent discussion, with full
references to previous proposals (by authors such as
Friedman and Torretti), is \cite{Mai98}.}.

We emphasise that this is not a mere problem of warring
intuitions---people agree about which theoretical
structures are intuitively fixed in these theories: for
example, that the spacetime metric is fixed in special
relativity, but not general relativity. The problem is to
frame a definition that captures the intuitions.

This problem is illustrated by the fact that this third
meaning is, for the most part, logically independent of the
first two. The exception is that the third meaning
presumably implies the second. Or in terms of the
contrapositive: being dynamical presumably implies being
not fixed in the third sense. For it would be an odd theory
that has such little variety in the configuration of matter
fields, from model to model, as to make such an influence
completely uniform, giving fixity in the third sense. But
the converse is false: a structure could be unfixed in the
third sense, but not dynamical---a sheer happenstance of
the world, varying from one model to another but
uninfluenced. An obvious example of this is the global
topological structure of time in Newtonian theories,
written as allowing topologies other than the real line.

And this third meaning is logically independent of the
first. Being classical does not imply being fixed in the
third sense: {\em cf.} the metric in general relativity.
And the converse implication also fails: for example, in
the standard quantum theory of a system moving in
3-dimensional Newtonian space, the algebraic form of the
canonical commutation relations $[\hat x_i,\hat
p_j]=i\delta_{ij}\hbar$ $i,j=1,2,3$, is fixed in the third
sense, but it is not classical since it involves operators.

We do not need to pursue more precise definitions of these
three meanings of fixity. But we should make three remarks,
in increasing order of importance.
\begin{itemize}
\item
First, we should note a trivial reason why fixity is vague,
which can cause confusion. It arises from the fact that,
even in a single theory, a word such as `time' or
`electromagnetic field' usually stands for a whole cluster
of mathematical structures, often hierarchically organized
from general (logically weaker) down to specific
(stronger). Typically, some general structures are fixed
(in all three of our senses) even if specific ones are not.
For example, the fact that the set of all time points has
the cardinality of the continuum is fixed (in all three
senses) in general relativity, though its metrical, and
global topological structure is not (in the second and
third senses). Similarly, the differentiable nature of the
electromagnetic field (away from point-particle charges) is
fixed (in all three senses) in classical electromagnetism.
So to say, for example, ``time is not fixed in general
relativity'' is somewhat ambiguous: we are really saying
that some aspect of time---contextually determined as being
of interest: usually, its metric structure---is not fixed.

\item
In discussing quantum gravity, one must beware of a
temptation to think that the third sense of fixity implies
the first: that being the same in all models implies being
classical. The temptation arises because it is common in
quantum gravity programmes that quantize a classical
theory, to take as fixed in the third sense, a structure
that is classical. For example, around 1970 there was an
(unsuccessful) programme to quantize general relativity by
quantizing metric perturbations around the fixed,
background spacetime metric of special relativity.

\item
Finally, we emphasise that the problem of precisely
defining the third sense of fixity is closely connected
with the diffeomorphism invariance of theories such as
general relativity, and thereby with the deep
interpretative issue of whether or not, and in what sense,
spacetime points are `real'. Here, `diffeomorphism
invariance' refers to the fact that the physical content of
the physical content is wholly unaffected by applying any
smooth, invertible transformation (called a
`diffeomorphism') between spacetime points, which thus
function as mere `pegs' on which to `hang' the fields; this
is the force of Einstein's famous `hole argument'. This
feature suggests that---quite apart from difficulties at
the scale of the Planck-length---spacetime points should
not be taken as real objects, even in interpreting
classical general relativity: rather they are an artefact
of the way we have formulated the theory. This leads to the
next Subsection.
\end{itemize}

\subsection{Spacetime and Matter}
\label{SubSec:Matter} In the introduction to quantum
gravity in Section \ref{Sec:EnterpriseQG}, we barely
mentioned something that everyday thought takes to be much
more unproblematically real than space and time: namely,
matter! And Section \ref{Sec:QGeom} will similarly say
little. Here, we will first discuss why matter comes to
play second fiddle to space and time; and then we will
briefly sound a warning against taking spacetime as in some
way more basic than matter.

 There are special technical reasons why so little was said about
matter in Section \ref{Sec:EnterpriseQG}; and also a
deeper, general reason. From a technical perspective, many
of the conceptual issues surrounding the emergence of time,
or the problem of time, are much the same, whether or not
one considers matter. For example, as Kucha\v{r} (this
volume) says, ``matter is not the heart of the problem [of
time in geometrodynamics]; it is gravity''.

The deeper, more general, reason arises from the way that
matter has come to be treated in twentieth-century physics.
Namely, some of the most distinctive characteristics of
matter, at least as idealized by the mechanical philosophy
of the seventeenth century---such as impenetrability and
continuity---turn out to be only apparent; and others, such
as massiveness, turn out to be shared with invisible fields
that extend in `empty space' arbitrarily far from their
sources. Further, matter itself has come to be treated as a
field, and this involves being represented mathematically
by an assignment of a mathematical object---at its
simplest: an item like a real number, or a vector, or a
matrix---to each point of space or spacetime.

Thus, with the rise of classical field theory, culminating
in general relativity, the field gradually became endowed
with properties previously thought peculiar to matter. Thus
the electromagnetic field was discovered to possess
mechanical properties like momentum and angular momentum,
and (after the advent of special relativity) mass-energy.
The `action-at-a-distance' of Newtonian gravity readily
admits a translation into the language of fields; and, of
course, the theory of general relativity was couched in
such language from the outset. Furthermore, this
assimilation of matter to the concept of a field continued
with quantum theory. Even elementary quantum theory has a
formulation (wave mechanics) which can be thought of as a
field theory. And this picture deepens with the advent of
quantum field theory: here physical quantities describing
matter are again represented by a field; albeit in a
transmogrified form in which an operator (rather than a
simple real number) is associated with each space, or
space-time, point.

This development has of course had as great an influence on
the philosophical understanding of space and time, as of
matter itself. In short, the effect has been for matter to
now play second fiddle to space and time. As these field
theories, both classical and quantum, are presented to us,
their basic objects seem to be just spacetime points. All
else---matter, fields, and even the metrical structure of
spacetime, the `chronogeometry' of the world---are
represented as mathematical structures defined on these
points; in particular, as just emphasised, physical
quantities for matter and fields are represented by
structures defined point by point. And so it is natural to
presume that all these are to be construed as properties of
(and relations between) points, or higher-order properties
and relations. In short, we arrive at the doctrine now
called `substantivalism': that spacetime points are genuine
objects, indeed are the basic objects of physical theory.

So much the worse, it seems, for the relationist tradition,
stemming from Leibniz through Berkeley and Mach, of denying
that space, time (and spacetime), and their parts, are
objects; and of conceiving them instead as a system of
spatiotemporal relations between bodies. For the bodies
have become diaphanous and omnipresent fields, which are
themselves complicated structures of properties and
relations among spacetime points.

But we think one should be wary, both of this kind of
argument for substantivalism, and of the doctrine itself.
Space permits only the briefest statement of our reasons:
we will mention two.

The first applies to any theory, such as general
relativity, that is diffeomorphism invariant; or in the
terminology of Section \ref{SubSec:Fixity}, to any theory
whose only fixed structure (in the third sense of `the same
in all models') is the local topological and differential
structure of a manifold---so that, intuitively, any map
preserving this structure ({\em i.e.}, any diffeomorphism)
preserves the content of the theory. As mentioned at the
end of Section 4.1, diffeomorphism invariance is widely
taken to indicate that spacetime points are indeed not
objects: their appearing to be so is an artefact of how we
formulate the theory. This reason exemplifies a general
point about objects or ontology. Namely: we should be wary
of taking as the basic objects of our ontology (according
to some theory) those items that are postulated as the
initial elements in a mathematical presentation of the
theory. For it might be just a happenstance of our
formulation of the theory that these objects `come first':
a happenstance avoided by another formulation that can be
agreed, or at least argued, to be better.

The second reason applies to any theory of physical
geometry (where `geometry' can be taken as including time
as well as space, as in `chronogeometry'). So it makes no
presuppositions about what structure is fixed in the sense
of Section \ref{SubSec:Fixity}. It has as its target, not
the doctrine that spacetime points are `real' but rather
the doctrine that the metric structure of spacetime is
intrinsic to it. In short: even supposing that spacetime
points (or regions) are real, the fact that our only
access, even in principle, to the metric structure of
spacetime is through the behaviour (perhaps ideal
behaviour) of matter---the proverbial rods and
clocks---suggests that metric structure is not intrinsic to
spacetime, but rather a relational matter reflecting the
nature both of spacetime and of matter.

This last suggestion is very vague, but we cannot here
develop it. It must suffice to mention one well-explored
way of doing so, namely the Machian theories of Barbour
{\em et al}. (For more details, see Barbour's article in
the present volume, especially Section 4.) In a nutshell:
these theories postulate
\begin{enumerate}
\item A space, often a relational one, {\em i.e.}, a space whose
geometry is fixed by the spatial relations between bodies;

\item A `least-action' principle on this space, which defines
both the dynamics of the system and the metric of time.
\end{enumerate}
Incidentally, this definition of the temporal metric
exemplifies the themes of Section 2 and 4.1. For this
metric is emergent in the sense that it is not a basic
postulate of the theory, and is also unfixed in the second
and third senses of Section \ref{SubSec:Fixity}. But it is
reduced in the same strong sense as the microcanonical
measure in Section 2.2.2: {\em i.e.}, it is explicitly
definable in a language sufficiently rich to express
calculus.

To conclude: we think these two reasons should make one
cautious about any quantum gravity programme that
postulates a spacetime manifold (whether fixed or unfixed,
in the senses of Section \ref{SubSec:Fixity}) that,
according to the theory, can be `bare' (matter-free). And
there are several such programmes, including the several
approaches to quantum gravity based on the ideas of
canonical quantization.

\subsection{Interpreting Quantum Theory}
\label{SubSec:InterpretQT} We turn to the last piece of
`stage-setting' needed for Section \ref{Sec:QGeom}: issues
concerning the interpretation of quantum theory. This is a
large and complex subject, but fortunately our discussion
can be very selective. Our topic of the emergence of time
means that we must be concerned primarily with whether
quantum theory has, or could have, the wherewithal to
describe the classical spatio-temporal realm around us.
Therefore we must take notice of the measurement problem,
with its implicit threat that physical quantities
pertaining to macroscopic objects take no definite value
(but only a probability distribution)---contrary to our
experience. More generally, we must be concerned with the
interpretation of the quantum state-vector. But there are
many interpretative issues about quantum theory, such as
entanglement and non-locality, which we need not engage
here .\footnote{Another sense of `describing the (unique)
classical realm' will occur in the discussion in Section
\ref{euclunique} of the no-boundary proposal: a sense
related to the idea of a theoretically motivated boundary
condition.}

In fact, we need to raise two issues. The first arises in
elementary quantum theory; the second has some aspects that
are specific to quantum gravity. They are, in order:
\begin{enumerate}
\item[(1)]
classical limits, approximations and decoherence in quantum
theory;

\item[(2)]
the interpretation of the quantum state; especially the
quantum state of the Universe (which will be taken in the
programmes focussed on in Section \ref{Sec:QGeom} as a
functional of 3-geometries).

\end{enumerate}

\subsubsection{Classical limits, approximations and decoherence in
quantum theory}

One of course expects the emergence of time (with `time'
taken in our usual sense of `classical spacetime') within a
quantum gravity programme to involve some sort (or sorts)
of classical limit of quantum theory, and/or some sort of
classical approximation methods within quantum theory. And
so it does, in our chosen quantum gravity programmes, as we
shall see in Section \ref{Sec:QGeom}. Indeed, we will see
there examples of each of the variety of approximations
discussed in Section 2.4: (a) the neglect of some
quantities; (b) the selection of a subset of states; (c)
combinations of (a) and (b), {\em i.e.}, a subset of states
approximating another theory as regards some quantities.

Here we just register that these various ways also occur
even in the context of elementary quantum theory; and that
the ideas and techniques used in this context do get
applied (often suitably adapted) by the quantum gravity
programmes discussed in Section \ref{Sec:QGeom}.
\begin{itemize}
\item
As to (a): we said in Section 2.4 that it was well-nigh
universal practice in physical theorizing. In elementary
quantum theory, (a) is also appealed to in foundational
arguments. And this occurs in various ways. First, various
approaches to the measurement problem argue that one need
only avoid macroscopic indefiniteness of value, for a
select subset of quantities. Second, in the theory of
decoherence (more details below), neglecting some
quantities (roughly, quantities defined on the environment
of the system one is concerned with) is the crucial point
in a physical argument that one avoids macroscopic
indefiniteness of value for some other quantities (defined
on the system).

\item
As to (b): when a probabilistic theory such as quantum
mechanics replaces a deterministic one such as classical
mechanics, one naturally hopes that the expectation values
of some quantities in the probabilistic theory might obey
the same equation of motion as the actual values obeyed in
the earlier theory. In general this is {\em not\/} so for
quantum mechanics; even for the basic quantity of classical
equations of motion, viz.\ position. But there are
circumstances in which it is so, or approximately so, for
some special states.

\item
As to (c): In some cases, expectation values for quantities
such as position, approximately obey the classical
equation, because special conditions relating {\em both\/}
to the states involved, and to a selection of quantities,
hold good.
\end{itemize}

We emphasise that, for the most part, these ideas and
techniques about classical limits and approximations are
not controversial. That is: they can be invoked in order to
address the measurement problem, or more generally issues
about the interpretation of the state-vector; but they do
not in themselves solve such problems---they at most
contribute to (the technical core of) a solution. In
particular, this applies to the use of these ideas and
techniques in quantum gravity: so nothing in the use made
of them in Section \ref{Sec:QGeom} will count as solving
these problems!

Similar remarks apply to another idea: decoherence. That is
to say: decoherence is an important aspect of quantum
theory's `recovery' of a classical realm; it has been
well-explored in the context of elementary quantum theory,
and has been invoked in the quantum gravity programmes in
Section 5. But we stress that it is not itself a solution
to the measurement problem.

Here we just need to register the main idea. Roughly
speaking, decoherence is the physical process of the
diffusion of coherence from a system to its environment;
here, `coherence' refers to the characteristic interference
terms of a quantum superposition, that distinguish the
superposition from a classically interpretable mixture of
states. In a bit more detail: decoherence is a process that
rapidly puts macroscopic objects (including tiny ones such
as dust particles) that interact with their environment
(even very slightly, such as with the microwave background
radiation) into mixtures, whose components are, in typical
cases, approximate position eigenstates. The process
diffuses the interference terms characteristic of the
macro-object's initial superposition into the environment,
so that its statistical behaviour is as if it is in a
mixture---and a mixture of states that have definite values
for familiar quantities like position, or quantities very
`close' to those like position. But we stress that this
mixture is `improper', {\em i.e.}, is not interpretable as
a matter of ignorance of which component is possessed. For
a recent review, cf. \cite{Giu96}.

\subsubsection{The meaning of a quantum state of the
Universe}\label{Subsubsec:MeaningQSU}

But we also need to say something about the interpretation
of the quantum state. This is not just a matter of it
always being salutary to emphasise the vulnerability of
quantum theory! Though most quantum gravity programmes put
more pressure on the formalism of general relativity, than
on the basic formalism of quantum theory, the endeavour of
quantum gravity {\em does\/} put pressure on the
interpretation of quantum theory---at least in the sense of
showing how debatable most interpretative tenets
are!\footnote{This pressure seems to us at least equal to
the interpretative pressure that quantum gravity put on
general relativity; for example, the trouble it cause for
substantivalism.} We will make two main points. The first
is specific to quantum gravity; the second is more general,
but will also be needed in Section 5.

(1) First, the endeavour of quantum gravity is closely
related to quantum cosmology, {\em i.e.}, the attempt to
have a quantum theory of the whole universe. And in quantum
cosmology, the traditional Copenhagen interpretation, with
its requirement of an observer external to the system, is
obviously inapplicable.

A related point is that among the variants of the
Copenhagen interpretation is one that requires, not an
observer, but a `classical background' external to the
system. And, arguably, one obvious candidate for such a
background is the continuous spacetime manifold of our
usual quantum theories (both quantum mechanics and quantum
field theories). But we admit to giving this sort of
interpretation little credence. For we see no good argument
for the necessity of such a background manifold as, for
example, (in Bohr's words) a `precondition of unambiguous
communication'.\footnote{Cf. the rejection in Section 3.4
of Kantian requirement for a manifold. No coincidence of
course, given the tenability of a Kantian reading of Bohr.}

(2) In general, the difficulty of interpreting the quantum
state makes it tempting to say something like (`N' for
naive):
\begin{equation}
\mbox{(N):\ }\int_{\Delta} \mid \Psi \mid^2 \mbox{ is the
probability for the the values of a quantity $A$ to {\em
be\/} in $\Delta$}
\end{equation}
where $\Delta$ is a subset of the spectrum of the
self-adjoint operator $\hat A$ (representing the physical
quantity $A$) to which the spectral theorem has been
applied to obtain a representation of vectors in the
Hilbert space as functions on $\hat A$'s spectrum.

    Indeed, it seems that (N) states the `core' of the
Born interpretation of the quantum state; and yet avoids
appealing to measurement, and also avoids (i) extra values
for quantities beyond the orthodox eigenvalue-eigenstate
link (`hidden variables'), and (ii) a physical collapse of
the quantum state. This temptation to assert the
proposition (N) is not specific to quantum gravity or
quantum cosmology: $\Psi$ need not be the state of the
universe.

But of course, (N) is hopelessly vague. First, it is clear
that if the quantum state does not collapse into having its
support confined to $\Delta$ (as (ii) requires), then the
phrase `the values \ldots to {\em be\/} in $\Delta$'
commits one to the extra values vetoed by (i). For what can
the phrase mean except that the quantity concerned takes a
value in $\Delta$ even though the support of $\Psi$ is not
confined to $\Delta$?

And once we accept that there are some such extra values,
we face various well-known hard questions.
\begin{enumerate}
\item Is there a preferred quantity or quantities $A$ which
get such values (rather than all quantities as suggested by
the formal schema (N))?

\item If so, can we choose the quantity or quantities so as to
help with the measurement problem, {\em i.e.}, so as to
secure a description of a classical realm?

\item There are also formal questions. Do the
quantities and values chosen avoid the traditional
no-hidden-variable theorems: both the algebraic theorems in
the tradition of von Neumann, and the non-locality theorems
in the tradition of Bell? And how do the values change over
time? That is: if we accept such values, we are obliged to
give a rule specifying how they evolve (deterministically
or stochastically) over time. And this rule must of course
mesh with our answers to questions 1.\ and 2. In
particular, if in answer to question 2.\ we secured a
classical realm at an instant, the rule for the evolution
of values should not lose it later!
\end{enumerate}

We need not consider how the various interpretations of
quantum theory (such as the pilot-wave and modal
interpretations) address these questions. But we should
briefly discuss the Everettian interpretation, since it has
been considered attractive in interpretative (and, even
more, in popular) discussions of quantum gravity.

The basic Everettian idea is familiar from the philosophy
of elementary quantum theory. It is that the particular
classical realm that is apparent to us corresponds to just
one component (summand) of the universal
state-vector---which always evolves deterministically,
never collapsing. As to the above questions, Everettians
traditionally tended to answer questions 1.\ and 2.\ by
proposing as the `preferred basis' in terms of which to
resolve the universal state-vector, the approximate
position eigenstates ({\em i.e.}, narrowly peaked
wave-functions) of measurement pointers: a proposal that
was widely, and we think rightly, criticized as giving too
fundamental a role to the notion of measurement; (for
example, \cite{Ken90}). But in recent years, they have
instead invoked the ubiquitous and very efficient process
of decoherence to give a principled, dynamically motivated,
specification of the preferred basis. We think this is an
improvement. In particular, it helps an Everettian explain
why the other classical realms he or she postulates as
corresponding to the other components of the universal
state-vector remain `hidden' from us: decoherence makes the
interference terms, that would reveal these realms, quickly
become far too small to be detected. \footnote{For more
discussion see, for example, \cite{Butt95} Section VI, and
\cite{Butt96} Sections 2, 3.}

As to question 3.\ above, Everettians traditionally tended
not to give a rule of evolution (even a stochastic one) for
the values of their preferred quantities. But recently they
have done so, often by adopting the `consistent histories'
formula for multi-time probabilities.\footnote{For more
discussion see, for example, \cite{Butt96} Section 5, 6.}

So much for the general problems of (N). If we apply (N) to
quantum geometrodynamics, in which $\Psi$ is a
wave-functional on 3-geometries, we get {\em yet more\/}
interpretative problems. We explain these in Section 5,
especially Section 5.3.

\section{Quantum Geometrodynamics}\label{Sec:QGeom}
\subsection{Prospectus; the Problem of Time}
So much by way of preliminaries! In the rest of this paper,
we will discuss how the idea of the `emergence' of time
appears in one particular approach to quantum gravity:
namely the canonical quantization programme, in the form
that leads to the famous Wheeler-DeWitt equation.

We shall discuss the emergence of time in two main
contexts, the first acting as a prerequisite of the
second.\footnote{For both contexts, we will be concerned
mainly with the emergence of local, rather than global,
aspects of space and time. For discussion in this volume of
the global structure of time, see the essays by Torretti
and Lucas.} The first, discussed in Sections 5.2---5.4, is
the Wheeler-DeWitt equation {\em per se\/}. Here, the
canonical quantisation of general relativity leads to a
quantum state-vector which is a functional of 3-geometries;
(this subject is known as `quantum geometrodynamics'). But
there are immense mathematical and conceptual difficulties
about this quantisation. Prominent among the conceptual
difficulties is what is what is perhaps the best-known
philosophical topic arising from quantum gravity: the
so-called `problem of time', discussed in detail in this
volume by Kucha\v{r}.\footnote{We heartily recommend
Kucha\v{r}, this volume, as a non-technical review. Among
the more technical reviews are \cite{Kuc92a,Ish93}; see
also \cite{BE99}. Our paper \cite{BI99} also discusses some
other aspects of quantum gravity from a philosophical
perspective.} We also will emphasise the problem of time,
since as we shall see it is closely linked to the emergence
of time.

In order to discuss the emergence of time in the face of
these various difficulties, our strategy needs must be,
after describing such a difficulty, to set it aside---if
only to raise another! More precisely, our plan, as regards
this first context, will be:
\begin{enumerate}
\item To introduce the problem of time in very general
terms (in this Subsection).

\item To introduce the canonical quantisation of general
relativity. This will include discussing (i) the origin and
status of the Wheeler-DeWitt equation; and (ii) the more
specific forms of the problem of time, and their relevance
to the emergence of time (Section 5.2).

\item To discuss the problem of interpreting a solution
of the Wheeler-DeWitt equation, supposing we could somehow
be given one (Section 5.3).

\item To discuss the idea, which arises from some
of the material in Sections 5.2 and 5.3, that time is an
approximate and state-dependent concept pertaining to a
semiclassical approximation to quantum geometrodynamics.
\end{enumerate}

In Section 5.5, we will turn to our second main context:
the so-called `Euclidean' programme, that aspires to
produce a particular solution of the Wheeler-DeWitt
equation using a functional-integral over (Riemannian)
metrics on a four-dimensional space. The goal is usually a
theory of quantum cosmogenesis---{\em i.e.}, a theory that
predicts a single, unique state for the quantum universe.

The material in Sections 5.2--5.4 will be a prerequisite
for this discussion. Indeed, there are two main connections
here. First, as we mentioned in Section 1: the Euclidean
programme's treatment of the emergence of time is beset by
the problem of time; and one of our goals is to correct the
impression in the popular literature about the Euclidean
programme (especially about the Hartle-Hawking no-boundary
proposal) that one can discuss the emergence of time
without having to address the problem of time. Second, more
technically: the treatment of the emergence of time in the
Euclidean programme relies on the treatment in Section 5.4
in terms of semiclassical approximations.

What then is the problem of time? As emphasized by
Kucha\v{r}, it is really a cluster of problems that arise
principally from the disparate ways in which time is
treated in quantum theory and in general relativity. The
main contrast goes back to the notion of fixity that we
discussed in Section \ref{SubSec:Fixity}: in quantum
theory, time is treated as a part of the fixed, theoretical
background structure, in {\em all three\/} of senses in
Section \ref{SubSec:Fixity}---classicality,
non-dynamicalness, and being the same in all models.

On the other hand, in general relativity---which is, after
all, a unified theory of space, time and gravity---time is
only fixed in the first sense of being classical. It is not
fixed in the third sense in so far as, for example, what
counts as a timelike vector depends on the space-time
geometry, and is hence model-dependent. And since this
geometry is the subject of dynamical laws, we can also say
that time is not fixed in the second sense. In short, time
is treated as an aspect of the system. Various conceptual
and technical difficulties in constructing a satisfactory
theory of quantum gravity can be traced, at least in part,
to this underlying difference in the `ingredient' theories.

The problem of time has been most studied in
geometrodynamics, {\em i.e.}, for the canonical
(Hamiltonian) approach to quantum gravity, using standard
geometric variables. Accordingly, Kucha\v{r} (this volume)
confines his discussion to this approach, and we shall
follow him in this. But we note that the problem of time
takes on a very different appearance (and is generally less
studied as such) in other approaches to quantum gravity
such as, for example, the superstring programme.

Finally, by way of prospectus: we should add that
Kucha\v{r} distinguishes three main forms of the problem of
time (as it appears in geometrodynamics), discussing them
in order. We will describe them in Section 5.2. But in
short, they are:
\begin{enumerate}
\item[(i)] the difficulty of finding an `internal time' in classical
general relativity;

\item[(ii)] the difficulty of finding a time `buried' in
the Wheeler-DeWitt equation; and

\item[(iii)] the difficulty of interpreting
the Wheeler-DeWitt equation in a fundamentally timeless
way, and accordingly treating time within our present-day
theories, general relativity and quantum theory, as an
approximate concept.
\end{enumerate}
Our discussion in Section 5.2 will follow the same order as
K\v{u}char; but since our topic is the emergence of time,
we will of course emphasise (iii).

\subsection{The Canonical Quantisation of General
Relativity} \label{Sec:CanQuGrav}
\subsubsection{Two types of canonical quantisation}
\label{SubSec:TwoTypesCanQu}

As we said in Section 3.3, the basic strategy of quantum
geometrodynamics is to cast general relativity in
Hamiltonian form, so that it describes the evolution in
time of the geometry of a 3-dimensional spacelike
hypersurface, $\Sigma$; and then to quantize this
Hamiltonian theory.

First, we should distinguish two different procedures, or
`recipes', for quantizing a Hamiltonian theory. Both are
called `canonical quantization', but the first (dating from
the 1920s) suits a Hamiltonian theory in which the number
of variables in the formalism equals the number of physical
degrees of freedom of the system described. For example,
the classical Hamiltonian mechanics of a point particle
moving in one dimension with position $x$, is quantized by
identifying a quantum state as a complex-valued function of
position $\psi(x)$, which is to evolve in time according to
the Schr\"{o}dinger equation (so that we write $\psi(x,t)$;
or, better, $\psi_t(x)$): more generally, any physical
variable $A$ becomes a linear operator $\hat{A}$ that acts
on the quantum states $\psi$.

The second type of canonical quantization (dating from the
1950s) suits a Hamiltonian theory in which there are more
variables in the formalism than there are physical degrees
of freedom; accordingly, the theory contains extra
equations that relate some variables to others. These
equations are called `constraints', and the theory (or the
system) is called `constrained'. One strategy for
quantizing such a system is to use the constraints to
eliminate some variables (this is called `solving the
constraints'), so as to get an unconstrained theory, which
is then quantized using the first procedure above. But an
alternative procedure is to quantize without solving the
constraints. Broadly speaking the idea is that given a
classical constraint equation, $c = 0$ say, where $c$ is
some function of the variables, one requires that the
quantum state $\psi$ obey $\hat{c}\psi = 0$, where the
`hat' indicates that the variables in $\hat{c}$ have become
operators. This is called `constrained
quantization'.\footnote{Strictly speaking, this procedure
is only correct for what are called `first-class'
constraints. We should also stress that there are other
procedures for quantizing a classical theory, the best
known of which---called `path-integral' quantization---was
developed principally by Feynman. It is more naturally
adapted to a Lagrangian, rather than Hamiltonian, form of a
classical dynamics; we shall return to it in Section 5.5.}

\subsubsection{The Wheeler-DeWitt Equation; its origin and status}
\label{SubSec:WDWEqn} It transpires that classical general
relativity in Hamiltonian form is a constrained system;
indeed the dynamics of the theory is coded in the
constraints (Kucha\v{r}'s `teorema egregium', equation (5)
of his paper in this volume). So one strategy for
quantizing the theory is to try to solve the constraints
before quantizing. This turns out to involve trying to find
a so-called `internal time' as a function of the canonical
variables of classical general relativity: a time which
could then serve as a time for the Schr\"{o}dinger equation
of the quantized theory. This is the most conservative
strategy from the quantum perspective since it manipulates
classical general relativity to get the kind of time that
is used in ordinary quantum theory. Unfortunately, however,
this cannot be done by simply taking over one of the
familiar `ordinary' times of classical general relativity
(for example, local proper time, or the global time
variables characteristic of certain simple cosmological
models). Rather, one has to confront the extremely
difficult problem of solving a collection of very
non-linear, coupled (elliptic) partial-differential
equations. So far, this has proved intractable. This, then,
is the first form of the problem of time: the problem of
finding a time `before quantisation'.

Another strategy is to try constrained quantization. Thus
one tries to quantize the Hamiltonian form of general
relativity, by postulating a complex-valued function of
3-geometry\footnote{The square bracket indicates that the
argument is itself a function (namely a specification of a
3-geometry on an entire spacelike slice). Such a `function
of a function' is usually called (by physicists) a
`functional'. The set of all 3-geometries (on a given
3-manifold $\Sigma$) is called `superspace'. It is a
infinite-dimensional space, since two 3-geometries can
differ from one another in infinitely many (independent)
ways. Sometimes, the word `superspace' is used for the
quotient set obtained by factoring out the action of the
spatial diffeomorphisms, associated with momentum
constraint discussed below. This quotient space is also
infinite-dimensional.} $\Psi[h]$ which will be subject to
appropriate operator constraints of the form $\hat
C\Psi=0$. Note that, by writing our prospective quantum
states as just functionals on superspace $\Psi[h]$, we have
ignored matter. To allow for matter fields, $\phi$ say, we
need to write a quantum state as $\Psi[h,\phi]$. However,
for simplicity we will ignore this qualification in what
follows, writing just $\Psi[h]$ {\em etc}---despite our
sympathy, expressed in Section \ref{SubSec:Matter}, for the
idea that matter is needed to make sense of geometry.

It transpires that general relativity has two types of
constraint, called the `momentum' and the `Hamiltonian'
constraints. When we quantize by requiring the `hatted
constraints' to send any state $\Psi[h]$ to $0$, we find
that the momentum constraint is readily interpreted: it
requires that $\Psi$ should take the same value at any pair
of 3-geometries $h$, $h'$ that differ only by a (smooth)
permutation of the points of the 3-dimensional space
$\Sigma$ on which they are defined. That sounds reasonable
since $h, h'$ seem physically equivalent, the points just
functioning as `pegs' on which to `hang' the
fields.\footnote{This is one of several points where the
problem of time is related to the diffeomorphism invariance
of general relativity, and Einstein's `hole argument'. As
discussed in Section 4.1, diffeomorphism invariance
suggests that spacetime points should not be taken as real
objects, even in interpreting classical general relativity.
In the canonical approach to quantum gravity, this argument
applies most immediately to points in space rather than
spacetime.}

But the quantum form of the Hamiltonian constraint is very
obscure. In the context of geometrodynamics, it is the
famous Wheeler-DeWitt equation which, for the sake of
reference, we give here as
\begin{eqnarray}
\lefteqn{
  -{\hbar^2\kappa^2\over 2}
   (\det h)^{-1/2}(x)\bigg(h_{ac}(x)h_{bd}(x)+
    h_{bc}(x)h_{ad}(x)-h_{ab}(x)h_{cd}(x)\bigg)
  {\delta^2\Psi[h]\over\delta h_{ab}(x)\,\delta
                h_{cd}(x)}-} \nonumber\\
&&\hspace{9cm}{(\det
h)^{1/2}(x)\over\kappa^2}\,R^{(3)}(x)\Psi[h]=0\hspace{1cm}
                                \label{WDE}
\end{eqnarray}
where $\kappa^2:=8\pi G/c^2$, and $R^{(3)}(x)$ is the
curvature scalar formed from the 3-metric $h$.

From the classical analogue, one would expect this equation
to express the dynamical content of the theory. Thus one
might expect $\Psi[h]$ to evolve according to some sort of
Schr\"{o}dinger equation. But the Wheeler-DeWitt equation
(\ref{WDE}) contains no explicit time parameter, so that to
make sense of the equation as describing evolution, one
apparently needs to find a time `buried' in it and its
associated formalism. And here we meet the problem of time
in its second form: it is very difficult to find such a
time `after quantisation'. This is the primary form taken
by the problem of time in geometrodynamics.

The difficulty of finding a buried time in the
Wheeler-DeWitt equation (and the related difficulty of
finding an `internal time' before quantisation) prompts the
idea that geometrodynamics, and perhaps quantum theory in
general, can---or even should---be understood in an
essentially `timeless' way. For if that were so, then one
would not expect time to appear in the basic postulates of
the theory: it would be an approximate concept, associated
perhaps with (i) some states (solutions) but not others; or
with (ii) some ranges of some variables but not other
ranges. This then is the third form of the problem of time:
to identify such an approximate concept, and argue that it
is adequate. Thus we arrive at our own topic: the emergence
of time.

Note however that the second and third forms of the problem
of time---finding a time `buried' in the Wheeler-DeWitt
equation, and finding some approximate notion of time in a
timeless geometrodynamics---should not be sharply
distinguished. In particular, the line of work we shall
discuss in Section \ref{SubSec:SemiClasQG}--5.5---the use
of ideas and techniques about semi-classical approximations
drawn from usual quantum theories---could arguably come
under either heading, since it involves extracting an
approximate notion of time from the Wheeler-DeWitt
equation.

But before discussing this line of work (in Sections 5.4
and 5.5), we must sound a note of warning. We must
emphasize the extreme interpretative difficulties that
surround not only this line of work, but also {\em all \/}
attempts to interpret solutions of the Wheeler-DeWitt
equation. Indeed, there are two points to be made here.
First, the difficulties of interpreting the quantum state
of the universe, discussed in Section
\ref{Subsubsec:MeaningQSU}, persist; and they become
entangled with the problem of time (Section 5.3).

Second, much of the discussion that follows is generically
deficient in the sense that no one who properly studies the
mathematics that lies behind the Wheeler-DeWitt equation
could seriously contemplate that the equation has
non-trivial solutions in any normal sense (other than in
gross approximations to the equation, such as that given by
using minisuperspace\footnote{Models that take account of
only a finite number of the dimensions of superspace are
called `minisuperspace' models. Though this sort of
restriction is radical, it need not involve neglecting all
but a tiny spatial patch of the 3-geometry; for a
sufficiently homogeneous 3-geometry might be specified by a
finite number of numbers. The usual, simplest example is
where the entire 3-geometry is coded by a single number:
the radius of a perfectly homogeneous and isotropic
universe.}): for example, from one perspective, the
notorious problem of the non-renormalisability of
perturbative quantum gravity rears its ugly head again at
this point. True: the main hope of the Ashtekar approach is
that it may eventually yield genuine solutions to its own
analogue of the Wheeler-DeWitt equation. However, this
analogue involve functions of spin-connections, or loops in
$3$-space, and hence the interpretation would be quite
different from that of geometrodynamics.

This situation inevitably prompts the sceptical question
why we are spending time and effort discussing possible
philosophical implications of an equation that is
mathematically meaningless! Are we as misguided as the
medieval scholastics are often taken to have been, in their
discussions of how many angels can dance on the head of a
pin? We believe not. Indeed, as philosophers know well, the
scholastics' discussions were not as misdirected as
folklore suggests. They addressed deep, maybe perennial,
issues about personal identity and spatiotemporal location,
in terms of their era's accepted ontology; which included
angels. We would like to think that our discussion stands
up as well.\footnote{Or at least, stands up as well to its
own era's scrutiny: as one might say---for them, the
Spanish Inquisition; for us, the Research Assessment
Exercise!}

\subsubsection{The problem of time vs. the emergence of time}
\label{3forms}

At this point we should emphasise that although the problem
of time and the emergence of time are closely connected in
various ways, they are distinct topics; even within
geometrodynamics. There are two main
differences.\footnote{The difference is clearer if we go
beyond geometrodynamics: Section \ref{Sec:EnterpriseQG}
makes it clear that the emergence of time can be discussed
apart from geometrodynamics---to which our formulations of
the problem of time are clearly confined.}

Perhaps the most obvious difference is that the problem of
time is mostly about time rather than space; while as we
said in Section \ref{Sec:Intro}, we intend `the emergence
of time' to also cover the emergence of spacetime, and so
space. At a technical level, the tendency not to dwell on a
`problem of space' stems from the fact that, as noted
above, in constrained quantization, the momentum constraint
---which, from a spacetime perspective, shuffles spacetime
points within a given 3-dimensional spacelike slice---is
much easier to deal with, conceptually and technically,
than the Hamiltonian constraint, which is related, at least
classically, to diffeomorphisms that map points from one
spacelike slice to another.

The second difference arises from the fact that the
emergence of time is about the emergence of {\em
classical\/} spacetime, simply because all our present
theories---whether classical general relativity or a
quantum theory of forces other than gravity---treat
spacetime classically. On the other hand, the problem of
time is about `finding' a time in {\em quantum\/} gravity
theories: and so not, in the first instance, about finding
some kind of classical limit of, or classical approximation
to, such theories. (A qualification: the third form of the
problem of time, above, is often focussed on finding such a
limit or approximation.) So the emergence of time is more
directly connected to questions about the classical limit
(and, more generally, interpretation) of quantum theory,
than is the problem of time (except in its third form). We
shall see this sort of connection in more detail in Section
\ref{SubSec:SemiClasQG}.

\subsection{Interpreting $\Psi$}
Suppose we are given a solution to the Wheeler-DeWitt: a
functional $\Psi$ of 3-geometries. How should it be
interpreted?

First of all, one must beware of a tempting error about the
relation between superspace and Lorentzian 4-manifolds
(spacetimes). It is tempting to think that a curve in
superspace determines a spacetime; and thus in cases where
Einstein's equations are satisfied, a model of general
relativity. But in general, this is not so: a curve in
superspace does {\em not\/} determine a spacetime, since
the curve does not dictate how its points, the
3-geometries, are embedded in the spacetime. Technically,
the latter information is coded in the `lapse function' and
the `shift vector': if these are given, then a spacetime is
determined. Conversely, a spacetime can be foliated in many
ways into a 1-parameter family of 3-dimensional slices each
with a 3-geometry. So in this sense, certain type of sheaf
of curves is equivalent to a spacetime. Note that, in some
simple (minisuperspace) models which neglect all but a
finite number of the infinite dimensions of superspace, the
lapse and shift are in effect fixed, and in this case a
curve in superspace {\em does\/} determines a spacetime.

With this warning in hand, let us press the question: How
should we interpret $\Psi$? One possibility might be to say
that $\Psi$ `predicts' a spacetime with a particular
4-geometry $g$ if (i) it has a non-zero value on any
3-geometry obtained by restricting $g$ to a spacelike
hypersurface of the spacetime; and (ii) any $3$-geometry
$h$ such that $\Psi[h]\neq 0$ arises as the metric induced
by $g$ on some 3-dimensional spacelike hypersurface.

However, such a view wholly neglects the way in which a
wave-function is used in standard quantum-theory to predict
probabilities. With a view to recovering the Born-rule, it
is tempting to return to the interpretation labelled (N) at
the end of Section 4.3, but now applied to superspace. That
is, it is tempting to say that
\begin{equation}
\mbox{(N):}\ \int_{\Delta} \mid \Psi[h] \mid ^2 {\cal D}h \
\mbox{is the probability for the 3-geometry $h$ to `be' in
the set $\Delta$.}
\end{equation}
This has been called the naive Schr\"odinger interpretation
\cite{UW89}: hence the `N'; but what are we to make of it?
Suppose we set aside all technical worries about the
definition of the measure symbolized by ${\cal D}h$.
Suppose also we set aside general worries about the meaning
of (presumably objective) probability for a state of the
whole universe.\footnote{Some modal metaphysicians are
happy enough to talk of such probabilities; for example,
\cite{Lew80}.} Still there is trouble, because we expect
theories to talk of probabilities of events {\em at a given
time\/}, while $\Psi$ and (N) make no reference to time.

To recover a time in the context of (N), it is natural to
hope that $\Psi$ will take a form that `contains' a time.
The obvious idea is that---setting aside the point above
that a curve in superspace does not uniquely determine a
spacetime---$\Psi$ should take non-negligible values only
on a single curve in superspace. If so, one might think of
$\Psi$ as `predicting' the corresponding spacetime. Or,
perhaps more realistically, one might hope that $\Psi$
takes non-negligible values only on a `strip' around a
curve in superspace; in which case one might hope to
identify some ensemble of curves all lying in this strip,
and even to use the values of $\mid \Psi[h] \mid ^2$ at
various points in the strip to talk about the probability
of evolving from one point to the other. And one might hope
to make out some such interpretation in the more general
case where $\Psi$ takes non-negligible values on a
criss-crossing set of strips.

But the trouble with schemes of this type is that there is
{\em no\/} reason to believe that, even if they existed,
solutions to the Wheeler-DeWitt equation would have
anything like these properties. Also, we must emphasise
that the qualification ``if they existed'' should be taken
very seriously: we recall our earlier caveats about the
mathematically ill-defined nature of the Wheeler-DeWitt
equation.

In addition to this particular reservation, a simple
variable-counting argument shows that the sense in which
$\Psi$ `contains' a time is quite different from, and more
subtle than, the above ideas about parameterizing curves in
superspace. In essence, it is a matter of time being a
degree of freedom (at every point in the 3-manifold
$\Sigma$) within the argument, $h$, of $\Psi$. More
precisely: to specify a 3-geometry on the space $\Sigma$
requires 6 real numbers at every point of $\Sigma$; three
of these correspond to making a choice of the otherwise
arbitrary coordinates on the $3$-space (this is closely
related to the spatial diffeomorphisms associated with the
momentum constraint), and two correspond to the physical
degrees of freedom of the gravitational field (the two
circular-polarization modes of a weak gravitational
wave)---but the last corresponds to the `internal time'
(`time before quantization'), mentioned at the start of
Section \ref{SubSec:WDWEqn}. This means, incidentally, that
in seeking a time within $\Psi$ (in accordance with the
`time after quantization' strategy), one is in fact thrown
back to the same kind of technical and conceptual problems
that beset the strategy of seeking a time before
quantizing. In short, we are back in murky waters!

\subsection{Semiclassical Approximations in Quantum
Geometrodynamics} \label{SubSec:SemiClasQG} We shall now
discuss the emergence of time in quantum geometrodynamics
in the approach that aspires to find time as an
approximate, semi-classical concept that arises `after
quantization'. (Since, as remarked above, time does {\em
not\/} emerge in the other main form of quantum
geometrodynamics (`time before quantization'), this
restriction is not as limiting as it might seem.)

The main idea is that (i) time should only emerge in the
context of some special states that are {\em approximate\/}
semiclassical solutions to the Wheeler-DeWitt equation; and
(ii) the time-variable that emerges depends on the state
chosen.

More precisely, one begins by considering a state of the
form $\Psi[h] = A[h]\exp(iS[h])$, subject to the conditions
that $S$ and $A$ are real, and $A$ is `slowly varying' in
comparison with $S$. Or, allowing for matter, one begins
with a similar form that incorporates matter variables
$\phi$. This ansatz is then substituted into the
Wheeler-DeWitt equation, and one seeks a solution as a
power-series in $S[h]$ (the so-called WKB expansion). The
leading-order contribution has the following significant
properties:
\begin{enumerate}
\item[(i)]
The exponent $S[h]$ obeys the same equation as does a
classical quantity (the `action') in general relativity
(viz.\ the Hamilton-Jacobi equation). Furthermore, this
connection between the phase of the quantum state and the
classical action is parallel to similar connections
established in standard quantum theories.

\item[(ii)]
{\em If\/} one can find a state-dependent (indeed,
$S$-dependent) `internal time', ${\cal T}$, obeying certain
conditions, then the entire lowest-order contribution to
$\Psi[h]$ (or allowing for matter, $\Psi[h,\phi]$), obeys a
Schr\"{o}dinger equation that uses a time-derivative with
respect to this time ${\cal T}$.
\end{enumerate}

The general character of the problems faced by this
approach can be summarized as follows. First, since one
starts with the Wheeler-DeWitt equation, one encounters the
problems mentioned earlier: in particular, the extreme
difficulty of giving the scheme even the semblance of a
proper mathematical meaning. In addition, it is extremely
difficult to find internal time variables except in the
simplest models---where, for example, the internal time
might be the radius of the universe. And, since the basic
interpretation of $\Psi$ is obscure, it remains unclear how
the fact that $S$ obeys the Hamilton-Jacobi equation of
general relativity is {\em meant\/} to yield classical
spacetimes.

But there are also problems that are distinctive of the
semiclassical approach itself. Some of these relate to
technicalities: for example, what should we make of the
fact that when the {\em next\/} term in the WKB expansion
is taken into account, the form of the Schr\"{o}dinger
equation is lost? But there are also purely conceptual
problems. Among the most interesting, from the viewpoint of
the emergence of time, is the question of how the
semiclassical approach proposes to recover a description of
the (apparently) {\em unique\/} classical spacetime around
us, despite the facts that in this approach:
\begin{itemize}
\item[(i)]
a single WKB state---{\em i.e.}, a single $S$---will
correspond to a {\em family\/} of classical spacetimes,
albeit all using the same internal time ${\cal T}$; (though
much is obscure in the recovery of classical spacetimes,
this one-many correspondence follows from a
variable-counting argument, and is not particularly
controversial);

\item[(ii)]
there seems no reason to exclude superpositions of WKB
states, each term in which gives rise to its own time
${\cal T}$ and its own family of spacetimes.
\end{itemize}

This last remark returns us to the topics of Everettian
interpretations, and decoherence, discussed in Section
\ref{SubSec:InterpretQT}. We shall not enter details about
the first of these, which is much discussed. Let us just
recall (from Section \ref{Subsubsec:MeaningQSU}) that the
Everettian denies that there is a sheer happenstance about
which `classical realm' is real, {\em i.e.}, about which
`branch' of the universal state is realized. Rather, the
Everettian proposes that all the realms (`branches') are
realized. In the present discussion, the Everettian can
presumably take a similar position. He or she can deny that
there is sheer happenstance about which classical
spacetime, among the family or families given by (i) or
(ii), is realized. Rather, all the spacetimes are realized.

But we should briefly discuss how the ideas and techniques
for modelling decoherence in usual quantum theories have
been adapted to semiclassical quantum geometrodynamics.
Recall (from Section 4.3.1) that decoherence is the
diffusion of coherence: it is the process whereby for a
system interacting with its environment, the interference
terms characteristic of a coherent superposition very
rapidly become far too small to be detected, leaving the
system in an (improper) mixture of states that are definite
for familiar quantities like position.

Of course, to adapt this notion to quantum geometrodynamics
(at least, as applied in the context of quantum cosmology),
one must allow for the fact that---if defined as the
totality of all that is---the universe has no environment!
In practice, studies of decoherence within quantum
geometrodynamics use some of the variables in the model
(either gravitational or matter variables) to act as an
environment for the others: these others constitute the
`system' whose state one wants to show to be a mixture.
Typically, inhomogeneous variables---{\em i.e.}, roughly,
spatially varying variables---are used as the environment
of the homogeneous modes. Thus one argues that the apparent
classical spacetime, with a spatially homogeneous geometry
and matter-distribution (on a large scale), is recovered;
at least in the sense that one obtains a mixture of states
(albeit an improper one) that are each approximately
definite for the homogeneous variables. So in particular, a
superposition of two or more WKB states (as in (ii) above)
that are defined only on the homogeneous modes will rapidly
evolve, to a very high degree of approximation, to a
corresponding mixture. And, as in Section
\ref{Subsubsec:MeaningQSU}, the Everettian who maintains
that there is a classical spacetime for each such state,
can appeal to decoherence to explain why the real
spacetimes other than our own are `hidden' from us: viz.,
decoherence makes the interference terms that would reveal
them far too small to be detected.

But we should also note two problems concerning decoherence
in semiclassical quantum geometrodynamics; (we set aside
general doubts about the contribution of the idea of
decoherence to solving the measurement problem in quantum
theory, for example based on the fact that the mixture one
obtains is improper). First, we should stress that the
tracing out of the `environment' variables to obtain a
mixture is ill-defined for reasons that, though technical,
are intimately tied to the conceptual problems about
interpreting the Wheeler-DeWitt
equation\cite{Kuc91a,Kuc92a}.

The second problem is more general, and more basic. There
is an obvious, major conceptual difference between
decoherence, as understood in usual quantum theories, and
as used here. In the former context, time is fixed in the
senses of Section \ref{SubSec:Fixity}, and decoherence is a
process that takes place in time. Here, on the other hand,
there is meant to be no time at the fundamental level; but
somehow, decoherence is meant to operate at this level to
make time emerge. So whatever `decoherence' may mean, it
cannot be construed as a temporal process.

So much by way of reviewing the emergence of time in
quantum geometrodynamics. We will see in the next Section
how the problems described in this review---the various
aspects of the problem of time, the murkiness of the
interpretation of a functional $\Psi$ on superspace, and
the problems of the semiclassical approach---all apply to
the Euclidean programme's approach to canonical quantum
gravity. Indeed, its treatment of the emergence of time
{\em relies\/} on the treatment given by semiclassical
quantum geometrodynamics.

\subsection{The Euclidean Functional Integral Approach to Canonical
Quantum Gravity} \label{SubSec:ECG}
\subsubsection{The general idea}\label{subsubsec:genidea}
The general idea is to construct wave-functionals on
3-geometries with the aid of a functional integral over
manifolds endowed with a Euclidean metric. To this general
idea, the no-boundary proposal adds a specification of
which types of manifold one should consider.

The origin of the `Euclidean' approach to quantum gravity
lies in Hawking's remarkable discovery in 1974 that a black
hole will radiate particles with a thermal spectrum via a
quantum-mechanical process \cite{Haw75}. Hawking's results
were quickly rederived using thermal Green's functions
which---in normal quantum field theory---are closely
connected with replacing time by an imaginary number whose
value is inversely proportional to the temperature. This
led Hawking to propose his `Euclidean' quantum gravity
programme in which the central role is played by
Riemannian, rather than Lorentzian, metrics (this being the
appropriate curved-space analogue of replacing time $t$
with $\sqrt{-1}t$). In particular, Hawking proposed to
study functional integrals of the form
\begin{equation}
    Z({\cal M}):=\int{\cal D}g\,
        e^{-{1\over\hbar}\int_{\cal M}|\det g|^{1/2}R^4(g)}
                            \label{Def:Z(M)}
\end{equation}
where the integral is over all Riemannian metrics $g$ on a
four-manifold $\cal M$ (and $R^4(g)$ is the curvature of
$g$, and det $g$ its determinant). The expression Eq.\
(\ref{Def:Z(M)}) generalizes naturally to include a type of
`quantum topology' in which each four-manifold $\cal M$
contributes with a weight $\chi({\cal M})$ in an expression
of the type
\begin{equation}
Z:=\sum_{\cal M}\chi({\cal M})Z({\cal M}).
\label{Z=sumZ(M)}
\end{equation}

It is not easy to give a rigorous mathematical meaning to
these objects but, nevertheless, the idea has been
extremely fertile. In particular, if applied to a manifold
with a single 3-boundary $\Sigma$, the expression Eq.\
(\ref{Z=sumZ(M)}) corresponds to a functional $\Psi[h]$ if
the functional integral is taken over all 4-metrics $g$ on
$\cal M$ that induce the given 3-metric $h$ on $\Sigma$.
Furthermore, the functional of $h$ thus defined satisfies
(at least, in a heuristic way!) the Wheeler-DeWitt equation
Eq.\ (\ref{WDE}); thus we might write (allowing also for
matter variables)
\begin{equation}
\Psi[h,\phi_0,\Sigma]=
 \sum_{\cal M}\chi({\cal M})\int {\cal D}\phi{\cal D}g\;
    e^{-{1\over\hbar}I(g,\phi,{\cal M})}
                        \label{path-int}
\end{equation}

This expression is the basis of the famous Hartle-Hawking
\cite{HH83} `wave-function of the universe' in quantum
cosmology. It is known as the `no-boundary' proposal, since
it is based on the idea that there is a sense in which the
universe has no `initial' boundary. To be precise: the
no-boundary proposal requires that the sum in
(\ref{path-int}) should be over compact manifolds $\cal M$
which have the (connected) three-manifold $\Sigma$ as their
{\em only\/} boundary; where $\Sigma$ is to be, as in
Section 5.2 onwards, the 3-manifold that carries the
3-geometries of quantum geometrodynamics.

We can now see the basis for the metaphor, much used in the
popular literature, that according to the no-boundary
proposal, the universe is created by `tunnelling from
nothing'. The word `nothing' just reflects (very
obscurely!) the idea that $\cal M$ has only $\Sigma$ as its
boundary. The word `tunnelling' refers to the facts that
(i) moving from a Lorentzian to a Riemannian manifold
corresponds, roughly, to moving from a time variable that
is a real number to one that is purely imaginary (in the
sense of complex numbers); and (ii) in normal quantum
theory, a good approximation to the probability of
tunnelling through a potential barrier, can be found by
computing the action $I$ for a solution to the {\em
classical\/} equations of motion with an {\em imaginary\/}
time; the probability amplitude in question is then
proportional to $\exp -I/\hbar$. We shall return to this
latter point shortly.

\subsubsection{The emergence of time in the Euclidean
programme} \label{clas-path-int}

It is clear that, like quantum geometrodynamics, the
functional integral approach makes fundamental use of a
manifold. This means not just that it uses mathematical
continua, such as the real numbers (to represent the values
of coordinates, or physical quantities); it also postulates
a 4-dimensional manifold $\cal M$ as an `arena for physical
events'. However, its treatment of this manifold is {\em
very\/} different from the treatment of spacetime in
general relativity in so far as it has a Euclidean, not
Lorentzian metric (which, apart from anything else, makes
the use of the word `event' distinctly problematic). Also,
we may wish to make a summation over different such
manifolds, as indicated in Eq. (\ref{path-int}). Finally,
as we shall discuss below, it is in general necessary to
consider {\em complex\/} metrics in the functional integral
(so that the `distance squared' between two spacetime
points can be a complex number), whereas classical general
relativity uses only real metrics.

Thus one might think that the manifold (or manifolds!) does
not (do not) deserve the name `spacetime'. But what is in a
name?! Let us in any case now ask how spacetime as
understood in present-day physics could emerge from the
above use of Riemannian manifolds $\cal M$, perhaps taken
together with other theoretical structures.

The main answer is what one would guess on the basis of the
remarks at the end of Section 5.5.1 about semiclassical
calculations of quantum amplitudes. That is: let us first
set aside the deep problems about the interpretation of
$\Psi$. Then the state-vector defined by Eq.\
(\ref{path-int}) will predict a classical spacetime (more
precisely: the correlations between position and momentum
variables that are characteristic of classical spacetime),
where it is well approximated\footnote{Strictly speaking,
this is what is true of normal quantum theory. However, the
possibility of making any proper mathematical sense of Eq.\
(\ref{path-int})---and hence of having a genuine
approximation to something that genuinely exists
mathematically---is extremely remote.} by the
semi-classical approximation to the functional integral.
Here, `the semi-classical approximation to the functional
integral' (also called `the saddle-point approximation')
means evaluating the integral by summing the contributions
given by the stationary points of the action.

In particular: if we choose to specify the boundary
conditions using the no-boundary proposal, this means that
we take only those saddle-points of the action as
contributors (to the semi-classical approximation of the
wave function) that correspond to solutions of the Einstein
field equations on a compact manifold $\cal M$ with a
single boundary $\Sigma$ and that induce the given values
$h$ and $\phi_0$ on $\Sigma$.

In this way, the question of whether the wave function
defined by the functional integral is well approximated by
this semi-classical approximation (and thus whether it
predicts classical spacetime) turns out to be a question of
choosing a contour of integration $C$ in the space of
complex spacetime metrics. For the approximation to be
valid, we must be able to distort the contour $C$ into a
steepest-descents contour that passes through one or more
of these stationary points and elsewhere follows a contour
along which $|e^{-I}|$ decreases as rapidly as possible
away from these stationary points. The wave function is
then given by:
\begin{equation}
    \Psi[h,\phi_0,\Sigma]\approx\sum_p e^{-I_p/\hbar},
\end{equation}
where $I_p$ are the stationary points of the action through
which the contour passes, corresponding to classical
solutions of the field equations satisfying the given
boundary conditions. Although in general the integral
defining the wave function will have many saddle-points,
typically there is only a small number of saddle-points
making the dominant contribution to the path integral:
(more precisely, that is what happens in normal quantum
theory where the mathematical expressions are reasonably
well defined).

For generic boundary conditions, no real Euclidean
solutions to the classical Einstein field equations exist.
Instead we have complex classical solutions, with a complex
action. This accords with the account in Section
\ref{SubSec:SemiClasQG} of the emergence of time via the
semiclassical limit in quantum geometrodynamics, in two
senses: the first rather negative (more precisely: `merely
permissive'), the second more positive.
\begin{enumerate}
\item
If the saddle-points of the integral {\em did\/} correspond
to real solutions of the Einstein field equations, with
real action $I$, the wave function would be of the
exponential form $e^{-I/\hbar}$ (note that the {\em sign\/}
of $I$ is not fixed). However, in order to predict
classical spacetime, a solution of the Wheeler-DeWitt
equation has to be of the form $e^{iS/\hbar}$, with the
phase $S$ satisfying the Lorentzian Hamilton-Jacobi
equation for general relativity. So, from that perspective,
it is just as well that the saddle-points give complex
classical solutions, with complex action: it means there is
some hope for recovering a classical spacetime.

\item
More positively, one can show---at least,
heuristically---that a WKB expansion of a state $\Psi$
arising from complex classical solutions (saddle points)
with complex action $I$ {\em does\/} predict classical
spacetimes (again: modulo all the difficulties adumbrated
earlier!)---given that a certain technical inequality holds
between the gradient of the imaginary part, $S$, of the
complex action $I$, and the gradient of the real part of
$I$. We note in passing that this exemplifies the idea
mentioned at the end of Section 2.4, of an approximation
scheme that combines in a single inequality the neglect of
some quantities and a constraint on some states.
\end{enumerate}

To sum up: if the relevant inequality holds (and if our
other approximating and interpretative assumptions hold
good!), then $S$ will be an {\em approximate} solution to
the Lorentzian Hamilton-Jacobi equation for general
relativity. The wave function will then be predominantly of
the form $e^{iS/\hbar}$, and in that case it {\em does}
define an ensemble of classical trajectories.

Of course, it must be emphasized once more that these
arguments are all very heuristic since the functional
integrals concerned do not exist in any proper mathematical
sense. The most that can really be said is that the
arguments have some validity in the context of
finite-dimensional minisuperspace models.

\subsubsection{The deceptive picture: Emergence is not a
process in time} In this Subsection, we spell out the fact
that the emergence of time, as treated in the Euclidean
programme, and in particular by the no-boundary proposal,
is {\em not\/} a process in time. Admittedly, this
corollary is probably evident in the light of the
discussion above. But it is worth emphasizing since a
considerable proportion of the popular, and philosophical,
literature about the Euclidean programme misses this point,
instead taking the Euclidean manifold to be somehow
`earlier' in a temporal sense than the classical spacetime
that emerges. This error results, of course, from ignoring
all the interpretative difficulties, approximation
assumptions {\em etc}.\ that we have been at pains to spell
out throughout Section \ref{Sec:QGeom}.

In the hands of some authors, this error has misleading
consequences. Namely, they argue that a treatment of the
emergence of time as a process in time faces conundrums,
even contradictions. (As one might guess, this is not hard
to argue: for example, if it is a process there are
presumably times before time emerged---surely a
contradiction.) As a result, the Euclidean programme is
falsely accused of facing various contradictions. Our moral
will of course be that while the Euclidean programme is
indeed very problematic---witness the discussion above---it
does not face knockdown objections from such
contradictions.\footnote{For these conundrums, see
\cite{SmiQ97,DelGuy96} and references therein. Lack of
space means that we cannot answer them {\em seriatim\/}
here. But note that this same wrong interpretation seems to
be behind Price's accusation \cite{Pri96} (pp.92--93) that
Hawking's explanation of the arrow of time on the basis of
the no-boundary proposal presupposes a temporal asymmetry;
namely by applying the no-boundary proposal only at one
temporal end, not both, of a temporally closed universe.
Yet it is clear from Section \ref{subsubsec:genidea} that
the no-boundary proposal is in no sense `applied at a
temporal end of a universe'; indeed, nor could it be. For a
philosophical review of the arrow of time in quantum
cosmology, we recommend \cite{Rid99}.}

It will be worthwhile to introduce our topic by relating
the ambiguity of the word `emergence', already noted in
Section \ref{Sec:Intro}---process in time, {\em versus\/} a
relation like reduction or approximation---to a distinction
between two scientific disciplines: quantum gravity, and
quantum cosmology. For the distinction is often ignored in
the the popular, and philosophical, literature about
quantum gravity.

 Though closely related in various ways, these disciplines are
different. In quantum gravity, one tries to reconcile
general relativity and quantum theory. As we mentioned in
Section \ref{Sec:EnterpriseQG}, this reconciliation can be
sought along many different avenues. But this endeavour
need not necessarily involve the idea of a theory of the
whole universe (let alone formulating it!). For example,
some research focusses on the physics of black holes---for
in the interior of a black hole, the very strong
gravitational field means that quantum gravity effects are
expected to be important. Other programmes, such as
superstring theory, concentrate on finding a role for
gravity as part of a unified theory of all the fundamental
forces---and theories of that type are as likely to be
concerned with the interactions between particles in an
accelerator as they are with the cosmos as a whole. On the
other hand, in quantum cosmology one seeks a
quantum-theoretical cosmology---by definition, a theory of
the whole universe.

The ambiguity in `emergence' is related to this
distinction. The relation arises from the fact that in
quantum cosmology, almost all work centers around theories
of the very early universe: in particular, with the hope of
finding a theoretical framework that avoids the
mathematical singularity that is inevitably present in the
account of Big Bang cosmology\footnote{Here, there is
indeed a link with quantum gravity: for, as in a black
hole, very strong gravitational fields will make quantum
gravity effects crucial in describing the Big Bang.} given
by classical general relativity. And here, quantum
cosmologists have made proposals that postulate a
`framework' (as we have seen it may well not deserve the
name `spacetime'!), which replaces the Big Bang singularity
of classical general relativistic cosmology, and from which
classical spacetime somehow `emerges'. Probably the
best-known example is the Hartle-Hawking no-boundary
proposal.\footnote{But there are many other proposals: one,
contemporaneous with the Hartle-Hawking proposal, and cast
in the language of geometrodynamics, is Vilenkin's
so-called `tunnelling' proposal. The original paper is
\cite{Vil88}; for a non-technical introduction see
\cite{Ish93b}; a non-technical introduction to the
competing Hartle-Hawking scheme is \cite{Ish88}.}

But this `emergence' is {\em not\/} a process in time:
Hartle and Hawking's proposed `framework', and others such
as Vilenkin's, stands in {\em no\/} temporal relation to
classical spacetime, or any of its parts (regions or
points), even very early ones.

This point is worth stressing, since three separate factors
tend to obscure it. First, the word `emergence' suggests a
process in time; in which case the framework (or perhaps
better: its parts) would presumably be earlier than
classical spacetime (or its parts). Second, the fact that
quantum cosmology focusses on the early universe makes it
very tempting to think of its proposals as concerning what
temporally precedes the epoch which our present theories
successfully describe. Third, many popular expositions of
some of the specific proposals, including the
Hartle-Hawking proposal, suggest the same (wrong!) idea.

Thus one often sees a picture in which a cone-like
spacetime structure (representing a cosmological solution
of classical general relativity) is attached to a spherical
shape that represents a Euclidean 4-manifold. This
erroneously suggests that the bottom sphere is
straightforwardly earlier than the classical cosmology
represented by the open cone in the top half of the figure.
%

But the 4-manifold is not earlier: there is no temporal
relation between the two halves represented in the figure
(or their parts)! Our discussion throughout Section 5 has
given several reasons why not. But let us rehearse some of
the more significant ones. In effect, they fall into two
groups: the first group specifically concerns the
no-boundary proposal; the second group returns us to the
interpretative difficulties in common between the Euclidean
programme and the quantum geometrodynamics of Sections
5.2--5.4---in particular, the cluster of problems called
`the problem of time'.

So first, recall that the no-boundary proposal involves a
{\em sum\/} over different manifolds $\cal M$. So there is
in any case no single bottom half-sphere, as the figure
suggests---just as there is no single trajectory followed
in a functional-integral approach to elementary wave
mechanics.

Second, the fact that a quantum tunnelling amplitude can be
given to a good approximation by $e^{-I/\hbar}$ where $I$
is the value of the action of a solution of the classical
dynamical equations with an imaginary time, does {\em
not\/} mean that this solution has any ontological status
in the quantum theory. The analogue of this point in the
Hartle-Hawking proposal is, of course, the facts that the
proposed 4-manifold has a Euclidean, not Lorentzian
metric---and that in general relativity, a Euclidean
spacetime has no more physical meaning than does an
imaginary-time trajectory in normal physics. In fact, the
interpretative difficulties become, if anything, even more
obscure if we use (as we must) a complex metric.

To this, one might well object that one can make sense of
the 4-manifold being earlier, in terms of the
topological-differential, rather than metric, structure
(for example, Section 3 of \cite{SmiQ97}). For example, one
could easily define (regardless of metric structure) curves
that start in the Lorentzian region of the manifold and
enter the Euclidean region. But here we meet the second
group of reasons, relating to such issues as the problem of
time. However, rather than listing some of those issues
again, we will venture a general statement why Hartle and
Hawking's 4-manifold has no temporal relations to classical
spacetime.

Namely, their entire proposal---the Euclidean manifold, and
other postulates described above---is intended as a way of
selecting a single wave-functional $\Psi[h]$ from the set
of all solutions to the Wheeler-DeWitt equation: (imagining
for the sake of the argument, that this equation is
well-defined).

This selection has various merits; (as well as various
problems, both conceptual and mathematical). But it does
{\em not\/} (and is not intended to) solve the problem of
time; nor does it give an account of the emergence of time.
That is: even if it overcame all its own problems, it would
give us nothing more (nor less!) than a wave-functional of
the universe obeying the Wheeler-DeWitt equation. The
problem of time outlined in Section
\ref{SubSec:WDWEqn}---here, principally the problem of
finding a time with which to make sense of the
Wheeler-DeWitt equation---would still have to be addressed.
So would the related issues of the emergence of time: here,
principally the issue, discussed in Sections 5.3--5.5.2, of
how to get a {\em classical\/} spacetime (a temporal
evolution of a {\em single\/} sequence of 3-geometries)
from this $\Psi[h]$.

To put the point in terms of the figure of a cone emerging
from a sphere: it deceptively suggests that the only
conceptual obstacle to joining a classical spacetime to the
Hartle-Hawking 4-manifold is the transition between the
Euclidean and Lorentzian metrics. But really, there are
several, subtly related obstacles: both the problem of
time, and the issue of the emergence of time, lie between
the two halves of the figure!\footnote{Though the
no-boundary proposal itself does not address these
obstacles, we emphasise that quantum cosmology as a
discipline has of course tried to do so.}

\subsubsection{Coda: Avoiding a boundary
condition?}\label{euclunique} Finally, we turn to an aspect
of the no-boundary proposal that does {\em not\/} fall
under our theme of the emergence of time, but which is of
such interest (and which has received such discussion) that
we must address it, albeit briefly.

We have in mind how, if at all, the no-boundary proposal
bears on the contrast between on the one hand, the laws of
a physical theory given by differential equations, or
equivalently their solution space; and on the other,
boundary conditions, which fix one solution in the
equations' solution space.\footnote{This contrast is of
course a special case of a broader contrast between the
laws of a theory, which even in physics are sometimes not
given by differential equations: for example, symmetry
principles; and statements of particular fact, which even
in physics are sometimes not given by numerical values of a
function, as boundary conditions are. We should also note
that a linguistic formulation of this contrast can break
down. For example, in the familiar case of fixing the
solution of a first-order ordinary differential equation by
specifying an `initial' boundary condition, the
differential equation plus boundary condition can be
combined in a single integral equation.} Interpretatively,
one takes the former to provide the description and
explanation of the possible patterns of behaviour of the
system in question; the latter determine, and `explain' the
individual case.

Laws and explanation are central controversial issues in
philosophy of science, which of course we cannot enter
here. But we only need two uncontroversial points. First,
there is a strong temptation to think of the laws (in our
special case: differential equations) as being truly
explanatory, while the statements of particular fact
(boundary conditions) are not explanatory, because they are
matters of `mere happenstance'. Second, this thought is
indeed questionable! After all, one can explain a
particular fact; and one can ask for an explanation of a
law. Agreed, the explanation of a particular fact will
typically invoke another such fact (usually an earlier one,
as in what philosophers call `causal explanation'), and one
can ask for an explanation of {\em that\/} fact---so that a
regress beckons, and one is tempted to think that one must
eventually accept a `mere happenstance'. But similarly,
when one can asks for an explanation of a law, the
explanation will typically invoke another such law
(relating them deductively, or by the one being a limiting
case of the other), and one can ask for an explanation of
{\em that\/} law---so again a regress beckons, and
apparently one must eventually accept a `mere happenstance'
of laws also.

These two points lead directly to how the no-boundary
proposal bears on the contrast between laws and boundary
conditions. In essence: the no-boundary proposal is
advocated as promising a stop to the regress of boundary
conditions, without having to settle on one as `mere
happenstance'---namely, by avoiding having a boundary
condition at all!

This proposal for a quantum cosmological wave-function that
avoids boundary conditions (in the traditional sense) in a
theoretically motivated way, is certainly very remarkable.
And we ourselves find it aesthetically pleasing. But we
needs must recall that it is not the only such proposal for
a `wave-function of the universe': Vilenkin's is the
best-known, and indeed contemporaneous, proposal
\cite{Vil88}. Of course, we cannot ask the no-boundary
proposal to itself rule out other such proposals, and in
that very strong sense `secure uniqueness'. But we should
note that at present, both these proposals and indeed
others are not ruled out; and in that sense, the science of
quantum cosmology as a whole has not avoided the `regress
of happenstance'. To do so would of course be a very tall
order: we do not mean this point as a criticism. Indeed, in
our opinion what matters is not the prospects for
eventually doing so, but rather the role of the no-boundary
proposal, and similar proposals, in understanding that
alluring idea---the nature, and emergence, of time.

\section{Conclusion}
\label{C}

To sum up our discussion: we have discussed two programmes
within quantum gravity, quantum geometrodynamics and the
Euclidean programme. We have focussed on the merits and
difficulties of their treatments of the emergence of time
(more precisely: the spacetime of general relativity).

 Broadly speaking, our conclusion is twofold. First, we concede that
there are large conceptual difficulties in both programmes'
attempts to have classical spacetime `emerge' (as well as
ferocious technical difficulties). But second, more
positively, we see no knock-down conceptual errors, or
philosophical howlers, in these attempts, as they have been
developed so far. In that sense, there are good prospects
for work for the future.

{\em Acknowledgements---} Chris Isham thanks the Mrs L.D.
Third Charitable Settlement for financial assistance during
during part of the course of this work. One of us (JB) is
very grateful to Katinka Ridderbos for discussions and
comments.


\end{document}